\documentclass[preprint2]{aastex}

\shorttitle{Catalog of Galactic $\beta$~Cephei Stars}
\shortauthors{A. Stankov \& G. Handler}

\begin{document}

\title{Catalog of Galactic $\beta$~Cephei Stars}

\author{Anamarija Stankov}
\affil{Research and Scientific Support Department, ESA-ESTEC, SCI-A, \\
2201 AZ Noordwijk, The Netherlands}
\email{astankov@rssd.esa.int}

\and
\author{Gerald Handler$^1$}\altaffiltext{1}{Erwin Schr\"{o}dinger R\"{u}ckkehrer fellow, Fonds zur F\"{o}rderung der Wissenschaftlichen Forschung, project R12-N02}
\affil{Institute for Astronomy, T\"urkenschanzstrasse 17, 1180 Vienna, Austria}
\email{handler@astro.univie.ac.at}

\begin{abstract}
We present an extensive and up-to-date catalog of Galactic $\beta$~Cephei
stars. This catalog is intended to give a comprehensive overview of
observational characteristics of all known $\beta$~Cephei stars, covering
information until June 2004. Ninety-three stars could be confirmed to be
$\beta$~Cephei stars. We use data from more than 250 papers published over
the last nearly 100 years, and we provide over 45 notes on individual
stars. For some stars we re-analyzed published data or conducted our own
analyses. Sixty-one stars were rejected from the final $\beta$~Cephei list, and 77
stars are suspected to be $\beta$~Cephei stars. A list of critically
selected pulsation frequencies for confirmed $\beta$~Cephei stars is also
presented.

We analyze the $\beta$~Cephei stars as a group, such as the distributions
of their spectral types, projected rotational velocities, radial
velocities, pulsation periods, and Galactic coordinates. We confirm that
the majority of the $\beta$~Cephei stars are multiperiodic pulsators. We
show that, besides two exceptions, the $\beta$~Cephei stars with high
pulsation amplitudes are slow rotators. Those higher amplitude stars have
angular rotational velocities in the same range as the high-amplitude
$\delta$ Scuti stars ($P_{\rm rot}
{\raise-.5ex\hbox{$\buildrel>\over\sim$}}$ 3\,d).

We construct a theoretical HR diagram that suggests that almost all 93
$\beta$~Cephei stars are main-sequence objects. We discuss the
observational boundaries of $\beta$~Cephei pulsation and the physical
parameters of the stars. We corroborate that the excited pulsation modes
are near to the radial fundamental mode in frequency and we show that the
mass distribution of the stars peaks at \linebreak 12 $M_{\sun}$. We point
out that the theoretical instability strip of the $\beta$~Cephei stars is
filled neither at the cool nor at the hot end and attempt to explain this
observation.
\end{abstract}

\keywords{stars: oscillations -- stars: early-type -- stars: fundamental parameters -- stars: interiors -- stars: Hertzsprung-Russell Diagram -- stars: variables: other}

\section{Introduction}

The past decade has seen many profound advances in our understanding of $\beta$~Cephei stars. The discovery of the $\kappa$-mechanism driving the pulsation of these stars \citep{MD92, DP93a} and the organization of many high-profile observing campaigns can be seen as recent highlights, and research into the physical properties of the $\beta$~Cephei stars has flourished in response. The number of known $\beta$~Cephei pulsators increases constantly, and recent years have seen us make several improvements to the way in which we discriminate between the many types of variable B-type stars. The exact definition of $\beta$~Cephei stars has itself been strongly debated over the years, and there is a good deal of ambiguity in most definitions. The recent advances in our understanding of $\beta$~Cepheids demand that a new refined definition be developed, and that a new $\beta$~Cepheid catalog be constructed and refined in line with this, examining and re-classifying all stars that have been previously identified or proposed as $\beta$~Cephei stars.

In recent years, two reviews on $\beta$~Cephei stars were published, describing the known group members from photometric and spectroscopic viewpoints, respectively. \cite{SJ93} published a review of all then-known $\beta$~Cephei stars including an extensive observational review of their astrophysical properties, and providing constraints on many of their key parameters. At the time, 59 $\beta$~Cephei stars had been identified. The following decade saw the identification of more than 40 new variables of this kind, bringing the total to almost 100, although the exact population has not been cataloged since the original 1993 review. In response to these new identifications, a complementary review paper was published investigating the spectral properties of bright $\beta$~Cephei stars that had detectable line profile-variations \citep{AD03}. Twenty-six objects could be examined in this way, allowing a better description of their physical properties and summarizing their pulsational behavior.

An excellent overview over $\beta$~Cephei and Slowly Pulsating B stars (SPB hereinafter) for which Geneva photometry is available is given by \cite{D02} in the form of an online catalog\footnote{http://www.ster.kuleuven.ac.be/$\sim$peter/Bstars/}. It provides the values of the Geneva indices as well as an homogeneous determination of stellar parameters based on calibrations of the Geneva system. This extensive compilation was one of the starting points for the present catalog.

Recent work has even demonstrated the presence of $\beta$~Cephei stars outside our own Galaxy \citep{PK02, KPS04} providing data for investigating this type of pulsation in objects of different metallicity (see Section 4.).

All of these achievements originated from ground-based observations.  
Today, at the dawn of the 21st century, asteroseismologists are preparing
to investigate variable stars from space, which will lead to the detection
of many more excited pulsation modes in these stars. The first step in
this direction was taken rather accidentally with the star camera of the
WIRE satellite \citep{BCL00} after the failure of its primary science
instrument. The first dedicated asteroseismology satellite, Canada's {\em MOST} \citep{MKW00, WMK03}, was launched successfully in 2003 June and is
returning valuable data on variable stars. Several other satellites
investigating stellar pulsation will soon follow.

In this paper, we attempt to refine our understanding of $\beta$~Cephei stars by cataloging the physical and pulsation properties of the entire confirmed population. This provides a comprehensive observational framework within which newly detected short-period pulsating stars can be classified. It will also be an aid for the classification of the vast amount of new pulsating stars that will be discovered in the near future.  Using this observational framework, we reevaluate the membership of every object that was once classified as, or suspected to be, a $\beta$~Cephei star.

Section 2 of this paper provides a brief description of the historical
classification of $\beta$~Cephei stars including information on
asteroseismic space missions. In Section~3 we describe different groups of
variable stars of spectral type B, from which we derive our working
definition of $\beta$~Cephei stars. Section 4 lists the properties that
have been examined for each of the stars in the catalog, explaining our
reasoning behind their use. Section 5 provides detailed analyses of the
entire data set from which we construct the observational framework, which
in turn is later used to aid the identification of $\beta$~Cephei stars.

In Section 6 we discuss the observational boundaries of $\beta$~Cephei pulsation. The conclusions and a definition of the $\beta$~Cephei stars are presented in Section 7. Tables of confirmed, candidate, and rejected $\beta$~Cephei stars can be found in Section 8 together with supplementary information on many of the objects.  There we also give lists of pulsational frequencies for all confirmed $\beta$~Cephei stars. In Section 9 we list all references and their corresponding numbers in the tables.

\section{A brief astrophysical history of $\beta$~Cephei stars}

The $\beta$~Cephei pulsators have been known to the astronomical community for more than 100 years. The variability of the prototype of this class of variable stars, $\beta$~Cephei, was discovered by \cite{F02}. A spectral analysis led him to the conclusion that this star ''is one of the Orion type ... , of which group the typical star is $\beta$~Canis Majoris.'' As a result, he named this group of stars the \linebreak $\beta$~Canis Majoris stars. At that time, the period of $\beta$~Cephei could not be determined with certainty. Some years later, the first radial velocity curve for this star was published by \cite{F06}. \cite{G13} discovered light variations of $\beta$~Cephei with the same period as the radial velocity variations, and it was also noted that the amplitude of the latter was not constant \citep[e.g.,][]{H18}. The correct interpretation for this phenomenon was found by \cite{L51}, who suggested that non radial pulsations are present in some $\beta$~Canis Majoris stars.

This group of stars comprised a rather wide range of variable B type stars and for many decades, at least up until the 1980s, these stars were known as $\beta$~Canis Majoris or $\beta$~Cephei stars. This redundant naming appears to have caused confusion among some authors. In, e.g. the {\em Hipparcos} catalog \citep{ESA97}, several stars are claimed to be classified as $\beta$~Cephei stars for the first time, but were actually already classified as such under their previous name of $\beta$~Canis Majoris stars. 

\citet{S77} discovered that some of these $\beta$~Canis Majoris stars were
spectroscopic variables and he called them the 53 Per stars. The term
Slowly Pulsating B (SPB) stars was introduced by \cite{W91} for
photometric B type variables pulsating in high radial order {\em g}-modes
(gravity modes) with periods in the order of days. The \linebreak 53 Per and SPB
stars contain stars that are pulsating with periods longer than the
fundamental radial mode and both groups have several members in common.
The separation between $\beta$~Cephei and SPB stars is a logical one based
upon the physical fact that the first group pulsates mainly in {\em p}-modes
(acoustic modes) of low radial order and the second in {\em g}-modes of high
radial order. This implies that their pulsational driving mechanisms
operate in zones of different thermal time-scale. The $\beta$~Cephei stars
usually have one or more periods similar to that of the fundamental radial
mode or the first non radial {\em p}-mode \citep{LA78}.

In their extensive review paper, \cite{LA78} defined the class of $\beta$~Cephei stars for the first time: ''{\em These stars have the same short period for their light variation and radial velocity variation.}''

The pulsational driving mechanism for the $\beta$~Cephei stars was unknown
for a long time. After a revision of the metal opacities \citep{IR87},
\cite{MD92} were able to compute models for $\beta$~Cephei stars in which
the fundamental radial mode became pulsationally unstable for
metallicities $Z > 0.03$.  They found that the size of the theoretical
instability strip for these stars depends on the abundance of heavy
elements and that the pulsation mechanism prefers low-frequency
oscillations. Only modes with a pulsation constant Q $>$ 0.032 became
unstable in these models. Furthermore, \cite{MD92} found that the
theoretical instability region is larger than the $\beta$~Cephei region in
the HRD and that the same pulsation mechanism could be present in luminous
blue variables (LBVs). 90 years after the first discovery of the pulsation
of $\beta$~Cephei, models could be calculated where $\beta$~Cephei type
pulsation was driven.

Refined computations by \cite{DP93a} and \cite{GS93} showed that pulsational instability could be reached for models with $Z > 0.02$. Instability was no longer restricted to the fundamental mode, but overtones were predicted to be pulsationally unstable as well. The current theoretical knowledge on the driving of $\beta$~Cephei pulsations has been summarized by \cite{P99}.

Many $\beta$~Cephei stars have been discovered to oscillate in several different pulsation modes \citep[e.g., see][]{J78}. This opens the possibility to explore the interior structure of these stars by using asteroseismology, i.e. deciphering their pulsational mode spectra and modeling them theoretically. Examples can be found in, e.g. \cite{DJ96,DJ99}. Indeed, \cite{ATD03} were able to understand the pulsational mode spectrum of the $\beta$~Cephei V836 Cen and to perform a first seismic analysis of the star. A recent multi site, multi technique campaign for another $\beta$~Cephei star, $\nu$ Eridani \citep{HSJ04,ADH04} enabled seismic modeling as well \citep{PHD04,AST04}.

The success of asteroseismic studies crucially depends on the detection and identification of as many modes of pulsation of the star under consideration as possible. Consequently, excellent measurement accuracy must be reached, which is best done from space.

{\em MOST} ({\em Microvariability and Oscillations of Stars}; \citealt{WMK03}) is
Canada's first space telescope and the very first dedicated
asteroseismology satellite delivering data. It will be followed by COROT
\citep{B03}, a French-led European mission with the goal to perform
asteroseismic observations as well as to detect exo planets transiting a
parent star.

All these developments in recent years led the authors of this work to the conclusion that it is time for an updated, homogeneous catalog of $\beta$~Cephei stars. This is not only useful for the target selection process for the upcoming space missions, but also important for the understanding of this group of pulsating stars as a whole.

\section{B-type variables, definition and selection of the $\beta$~Cephei stars}

As already mentioned, the $\beta$~Cephei pulsators are generally considered to be early-type B stars (B0$-$B2.5) with light and radial velocity variations on time-scales of several hours. As they are not the only variables of spectral type B, it is important to delineate what separates them from other variables. For instance, the SPB stars are later type B stars (B2$-$B9)  with light, radial velocity, and line profile variability with periods of the order of a few days \citep{DA02}.

The Be stars are defined as non supergiant B stars having shown Balmer line emission at least once \citep{Co87}. They span the whole $\beta$~Cephei and SPB instability regions and stretch from late O-type to early A-type stars. The hotter Be stars of spectral types B6e and earlier can show light, radial velocity, and line profile variations. Be stars that vary periodically \citep[see][]{RBS03} are sometimes also called $\lambda$~Eri stars. Some Be stars can also show additional $\beta$~Cephei pulsations (see the discussion below).

$\zeta$ Ophiuchi stars are OB-type variables that show bumps moving through their line profiles, which may be caused by high-degree non radial pulsation \citep{BD99}. 

Some of the S Doradus stars or Luminous Blue Variables \citep[see,
e.g.,][]{VG01}, also have spectral types of or near B.

The chemically peculiar Bp stars can also show light and line profile variations \citep[see, e.g.,][]{BAL04}, and ellipsoidal variables may be present amongst variable B-type stars as well.

Two rather recently discovered classes of pulsating B stars are the short-period subdwarf B variables (sdBVs), also known as EC 14026 stars \citep{KKO97}, and the long-period sdBV stars \citep{GFR03}. The periods of the short-period sdBVs range between 2 and 9 minutes, and those of the long-period sdBVs are around 1~hr. Finally, the pulsating DB white dwarf stars also need to be mentioned.

Consequently, it is not easy to classify variable B-type stars correctly, in particular as some overlap between the different groups of variable stars occurs. For instance, $\beta$~Cephei itself is also a Be and a Bp star \citep[see][for a summary]{HH96}.

We therefore suggest the following definition of the $\beta$~Cephei variables: {\it The $\beta$~Cephei stars are massive non-supergiant variable stars with spectral type O or B whose light, radial velocity and/or line profile variations are caused by low-order pressure and gravity mode pulsations.}

Our choice of this definition was motivated by several reasons. In our
view, the main feature on which the classification of a pulsating star is based 
should be its pulsational behavior. For instance, any class of pulsating
stars should be known to be driven by the same self-excitation mechanism,
and their pulsational time-scales should be different and separable from
those of other types of pulsators. Of course, a particular locus in the HR
diagram could also assist, and sometimes be incorporated, in the
definition of a class of pulsating star.

Since the observational extent of the instability strip of the
$\beta$~Cephei stars may still not be accurately known (cf. Sect.\,6), we
did not want to limit our definition to a narrow range of spectral types.  
In addition, we do not take the existence of radial pulsation to be a
prerequisite for an object to be classified as a $\beta$~Cephei star,
because this would require a firm observational mode identification, which
is in most cases not available. By dropping this criterion that has sometimes been used in the past, we make the pulsation constant our main
criterion for classification. We also take into consideration that
many $\beta$~Cephei stars have been shown to pulsate both in radial and
nonradial modes, or any subset of these.

To apply our definition to the stars under consideration, we must link it to observables. Consequently, we consider an object to be consistent with our definition of a $\beta$~Cephei star in practice, if it shows convincing evidence for more than one variability period too short to be consistent with rotational or binarity effects, as checked by estimating the pulsation ``constant'' $Q$. Stars with only a single period were accepted if proof of the pulsational nature of the variations was found, such as color or radial velocity to light amplitude ratios typical of pulsation or variability (with, again, the period too short to be accounted for by other effects) present in more than one observable.

\section{Description of the catalog}

We list all objects that have to our knowledge ever been claimed to be
$\beta$~Cephei stars or candidates up to 2004 June. We selected them by an
extensive search in the literature and in data-bases (such as SIMBAD),
with the aim that we could collect all possible candidates. For all of
them, thorough bibliographic studies were performed to investigate the
latest findings on their nature. Where the data in the literature were
insufficient or inconclusive, we reanalyzed some of the measurements or
reevaluated the available information on these stars. We also performed
frequency analyses of the {\em Hipparcos} photometry \citep{ESA97} whenever
possible to assist with the classification of the variables. We note that because of aliasing problems in the {\em Hipparcos} data we did not attempt to determine individual periods but mainly used them to check the time-scales of the observed variability. Owing to the particular variability time-scales involved, aliasing was therefore not a problem for our purposes.

The SIMBAD database initially prompted us with 128 $\beta$~Cephei 
stars. Their classification often originated from the General Catalogue of 
Variable Stars \citep{KKP71} and subsequent name lists. We could confirm 66 of these and placed the others either on a list of candidate or rejected $\beta$~Cephei stars. The other objects in this catalog were selected from our literature searches.

We then scrutinized the literature on all these stars and checked whether they were consistent with our working definition of a $\beta$~Cephei star. We designate objects that have been claimed as $\beta$~Cephei stars, but where the observational evidence for their membership to the group is not fully conclusive owing to, e.g. poor or few data, as {\it candidate $\beta$~Cephei stars}. Some of these objects will indeed be $\beta$~Cephei pulsators, whereas others were added to this list because of the lack of evidence that they are not. In any case, all of these objects deserve more observational attention.

Stars that were claimed to be $\beta$~Cephei variables, but where we found evidence that they are not, are called {\it rejected $\beta$~Cephei candidates}. These are objects with variability time scales inconsistent with low-order {\em p}- or {\em g}-mode pulsation, objects whose claimed variability was disproved by subsequent or more extensive studies, or stars proven to vary because of effects other than $\beta$~Cephei pulsation, etc. This list also includes stars that were rejected by other authors in order to give a complete overview over all stars that at some point had been considered to be $\beta$~Cephei stars. We have also scrutinized the list of $\zeta$ Oph stars by \cite{BD99} as several of these variables have observed periods in the $\beta$~Cephei range. We found that the periods {\it in the co rotating frame} are consistent with $\beta$~Cephei pulsation for only three stars, which we include in this catalog.

The importance of Tables~2 and 3 is, besides their relevance for the description of the $\beta$~Cephei stars as a group, that they provide completeness of the catalog and that it can be traced how the less convincing candidates were judged by us.

We have not included $\beta$~Cephei stars or candidates outside our Galaxy into this catalog because detailed lists of the objects reported by \cite{KPS04} have not yet been published. For reasons of consistency we thus also exclude the LMC $\beta$~Cephei candidates by \cite{PK02} and \cite{SJ88}.

In Table~1 we present the complete list of Galactic $\beta$~Cephei stars. Table~2 contains the candidate $\beta$~Cephei stars, and Table~3 lists the former candidates that are not considered $\beta$~Cephei stars.  In the following, we describe the contents of Table~1. Tables~2 and 3 only contain part of this information.\\

1. {\em Identifiers.--} The first two columns of the catalog contain the identifiers of the stars, HD number, HR number or cluster identification, {\em Hipparcos} numbers, and Durchmusterung (DM) numbers. As DM number for the Southern stars we used the Cordoba Durchmusterung (CD) numbers, not the Cape Photometric (CpD) numbers! Some objects do have CpD numbers, and no CD numbers, such as V1032~Sco.

The third column lists different names given to these stars, such as Bayer or Flamsteed numbers, or variable designations according to the General Catalogue of Variable Stars.\\

2. {\em Coordinates.--} Right ascension and declination (top rows of cols. [4] and [5]) are given with epoch 2000. If inaccurate coordinates were found in the databases, we used a finder chart and matched it to the Digital Sky Survey plates, thereby determining the coordinates to a precision of $\sim$2".\\

3. {\em Pulsation period.--} The period of pulsation is given in days in column (6). If a star is multiperiodic, the period of the pulsation mode with the highest amplitude is given and an asterisk ($\ast$) next to the period indicates the multi periodicity. We apply the term "mono periodic" to stars for which only one pulsation frequency was found {\em up to the current detection limit}. If pulsation was detected photometrically and spectroscopically, we give the photometrically determined period.\\

4. {\em Amplitude.--} It is difficult to specify a unique amplitude of a frequency for a variable star measured in different photometric passbands and/or in radial velocity. This applies in particular for the cases of multiperiodicity and amplitude variability. Therefore, we chose the following approach: for all stars with a resolved pulsation spectrum, we list photometric peak-to-peak amplitudes of the strongest pulsation mode. If a star shows amplitude variability caused by beating of unresolved frequencies, we adopt the average peak-to-peak amplitude, and in case of mild amplitude variability of individual modes we adopt the average peak-to-peak amplitude of the strongest mode. For the two stars with the strongest amplitude variations (Spica and 16 Lac) we list no amplitudes. Finally, no amplitude is given for the three stars where only spectroscopic variability was detected.  This is denoted as ''n/a'' in the amplitude column.
  
We list Johnson $V$ amplitudes whenever possible. If data from this filter were not available, we used Str\"omgren $y$, Geneva $V$ (denoted $V_G$ in Table~1), or Walraven $V$ (denoted V$_{\rm W}$). In the latter case the logarithmic intensity amplitudes were multiplied by 2.5 to give magnitude units. All these filters have very similar effective wavelengths resulting in directly comparable amplitude values. Some stars have not been observed in these filters. In such cases we chose in order of decreasing preference Str\"omgren $b$, Johnson/Cousins $R$, Johnson $B$, and Cousins $I$. The use of data obtained in different filters was only applicable because the pulsation amplitudes of $\beta$~Cephei stars are very similar in the wavelength range spanned by \linebreak$B$ to $I$.\\

5. {\em Apparent magnitude and spectral classification.--} The next two columns give the apparent magnitude in Johnson $V$ and the spectral type according to the Morgan and Keenan system (MK). The $V$ magnitudes are taken from The General Catalog of Photometric Data (GCPD hereinafter)\footnote{http://obswww.unige.ch/gcpd/gcpd.html} by \cite{MMH97}. The spectral types are taken from the SIMBAD data-base.\\

6. {\em Rotational and radial velocities.--} The projected rotational velocity ($v\sin i$) is given in the next column.  If disagreements between different values in the literature were detected, either the more reliable source is quoted here or, if no distinction in quality of the data could be made, the lower value is given. Concerning the radial velocity (RV), the best or mean values are quoted, in an attempt to average out the RV variations over the pulsation cycle. The values for $v\sin i$ and RV are taken from various catalogs of radial velocities, or in some cases from original publications.\\

7. {\em Color indices from Str\"omgren photometry.--} The bottom row of columns 4-7 contains the Str\"omgren color indices $(b-y)$, $m_1$, $c_1$, and $\beta$. These data were obtained from the GCPD. These indices are used in preference to the Geneva colors, which are available for roughly the same number of stars. The Str\"omgren filters are more widely available and according to our experience, the combination of measurement accuracy and its conversion to theoretical parameters such as effective temperature and surface gravity via calibrations of color photometry favors Str\"omgren photometry in terms of achieving better accuracy in the derived basic stellar parameters. In addition, only the $c_1$ index may show some variation over the pulsation cycle of a $\beta$~Cephei star. We therefore find that the color indices we list are good representations of the mean colors of a star through its pulsation cycle.

For a detailed list of Geneva colors for many $\beta$~Cephei stars we refer to \cite{D02}\footnote{http://www.ster.kuleuven.ac.be/$\sim$peter/Bstars/}.\\

8. {\em References and notes.--} The references given in the last column are numbered and listed in Section~9. Short individual notes to several stars can be found in the table, whereas longer discussions of some objects are given at the bottom of each table. As references, we list selected papers that are directly related to the stellar pulsations or those which give useful additional information.  These would typically be discovery papers, those that reported most about the pulsations, or did further analyses such as mode identifications. We give no more than six references per star. \\

\subsection{Table of frequencies}

In Table~4 we present a list of frequencies for the stars from Table~1.  We refrain from listing all the claimed frequencies for all stars because the data are qualitatively inhomogeneous and some sources may not be reliable. The choice of frequencies listed originates from critical evaluations of literature data.

We consider a photometrically detected frequency as also spectroscopically detected if the variation is present and clearly recognizable in radial velocity analyses or line profile variations, but do not insist on detections in both of these spectroscopic observables.

For several stars, some frequencies reached detectable amplitudes only during some observations. We list all frequencies ever detected from analyses that convinced us.

\section{Analysis}

\subsection{Basic observational quantities}

\begin{figure}[ht]
 \vspace{7cm}
 \includegraphics{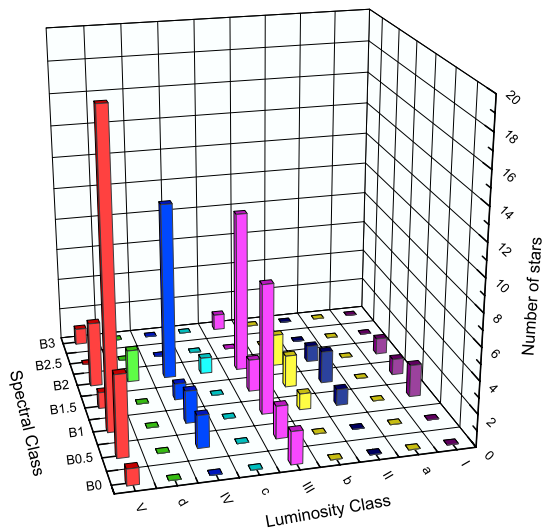}
\caption{Distribution of stars according to spectral type and luminosity class. The letters a, b, c, and d correspond to the intermediate luminosity classes I-II, II-III, III-IV, and IV-V.}
\label{3-D}
\end{figure}

In this section we present analyses performed on the intrinsic 93
$\beta$~Cephei variables. We analyze the distribution of spectral type
(see Fig.~\ref{3-D}), radial velocity (RV), projected rotational velocity
($v\sin i$), apparent brightness in Johnson $V$ and pulsation period ($P$)
(see Fig.~\ref{histo}). In addition, we examine the Galactic distribution
(see Fig.~\ref{gal}) as well as the dependence of the pulsational
amplitudes on the projected rotational velocities (see Fig.~\ref{Avsini}),
and thereby describe the $\beta$~Cephei stars as a group.

\subsubsection{Spectral type and luminosity class}

The three-dimensional histogram in Fig.~\ref{3-D}, which is inspired by Fig. 4 of Sterken \& \linebreak Jerzykiewicz (1993), shows the distribution of the confirmed 93 $\beta$~Cephei stars according to their spectral type and luminosity class. It shows that $\approx$20\% of the $\beta$~Cephei stars appear to be B1 dwarfs. A total of 66\% of the stars are of spectral type B1 and B2 and luminosity classes III-V. This distribution resembles very closely the spectral type range occupied by the confirmed $\beta$~Cephei stars from \cite{SJ93}, where almost all stars lie within B0 and B2.5. Most of the class V variables are members of open clusters (80\%). Two of the stars from Tab.~1 do not yet have a spectral type assigned (NGC 6910\,27 and V2187 Cyg) and for 3 stars no luminosity class was associated to the spectral type (NGC~663\,4, NGC 6910\,16, and HN Aqr). As will be shown in Sect. 5.2, the assignment of luminosity classes I--III to some of these stars must be erroneous.

\subsubsection{Projected rotational velocity}

The range of projected rotational velocity, $v \sin i$, extends from 0 to 300\,km\,s$^{-1}$ with \linebreak HD 165174 as the fastest rotator with 300\,km\,s$^{-1}$, closely followed by NGC 4755\,I with 296\,km\,s$^{-1}$. HD 165174 is also a Be star, whereas NGC 4755\,I went through phases where its pulsations were clearly detectable, but at other times did not reach a detectable level. Most $\beta$~Cephei stars seem to be rather slow
rotators (average $v \sin i \sim$100\,km\,s$^{-1}$), although this could in
part be due to a selection effect as the highest-amplitude pulsators are
slowly rotating stars. Hence, their variability is more easily detectable
and observable.

\subsubsection{Radial velocity}

\begin{figure}[h]
 \vspace{11.5cm}
 \includegraphics{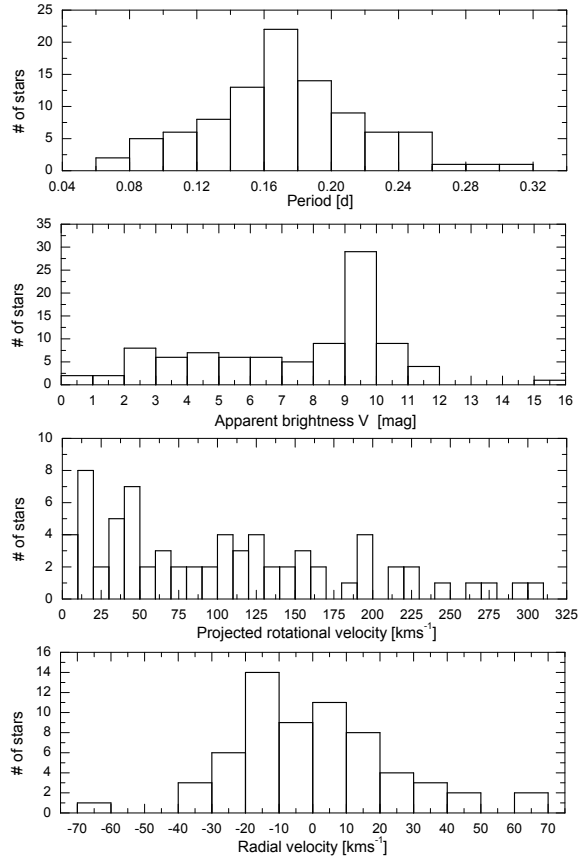}
\caption{Histograms of radial velocity, projected rotational velocity, apparent magnitude, and pulsation period.}
\label{histo}
\end{figure}

The radial velocities (RV) of the $\beta$~Cephei stars, as seen in Fig.~\ref{histo}, bottom panel, appear to be centered around $-$10\,km\,s$^{-1}$ but stretch up to +65\,km\,s$^{-1}$. This distribution is that of an average young Galactic disk population, which is not surprising.

\subsubsection{Apparent brightness}

The apparent brightness has a maximum at $V \sim 9.5$\,mag with 31\% of the stars; these are mostly cluster $\beta$~Cephei stars. The range of apparent brightness is between 0.6\,mag $ < V < $15.4\,mag, with $\beta$~Cen, $\alpha$~Vir, $\beta$~Cru, and $\lambda$~Sco as brightest stars with $V$ between 0.6\,mag and 2.0\,mag. The faintest stars with $V$ of 11.9\,mag and 15.4\,mag are HN~Aqr and V2187~Cyg, respectively. This information can be relevant for planning observational projects, and can be compared directly to Fig.\,3 of \cite{RB01}.

\subsubsection{Pulsation period}

The distribution of the pulsation periods has a peak at $\sim$0.17\,d, corresponding to 4\,hr. The shortest period is 0.0667\,d for $\omega^1$~Sco, the next shortest period is from Braes~929 with 0.0671\,d.  The two longest periods are 0.319\,d for Oo~2299 and 0.2907\,d for HD~165174.

Hence, we find that the observed range of periods for $\beta$~Cephei stars is between 0.0667\,d and 0.319\,d or 1.60\,h and 7.66\,h. The median of all periods is 0.171\,d.

Three of the confirmed $\beta$~Cephei stars show, so far, variability only in their line profiles. They are nevertheless included in the group of $\beta$~Cephei stars because they exhibit the same basic behavior as the {\it classical} $\beta$~Cephei stars. The lack of confirmation of their variability from photometric techniques is due to modes of high-degree $\ell$ in those stars, which are difficult to detect in photometric observations. We have only retained objects in Table~1 if their corotating variability period was consistent with $\beta$~Cephei pulsation.

\subsubsection{Galactic distribution}

\begin{figure}[h]
 \vspace{5.5cm}
 \includegraphics{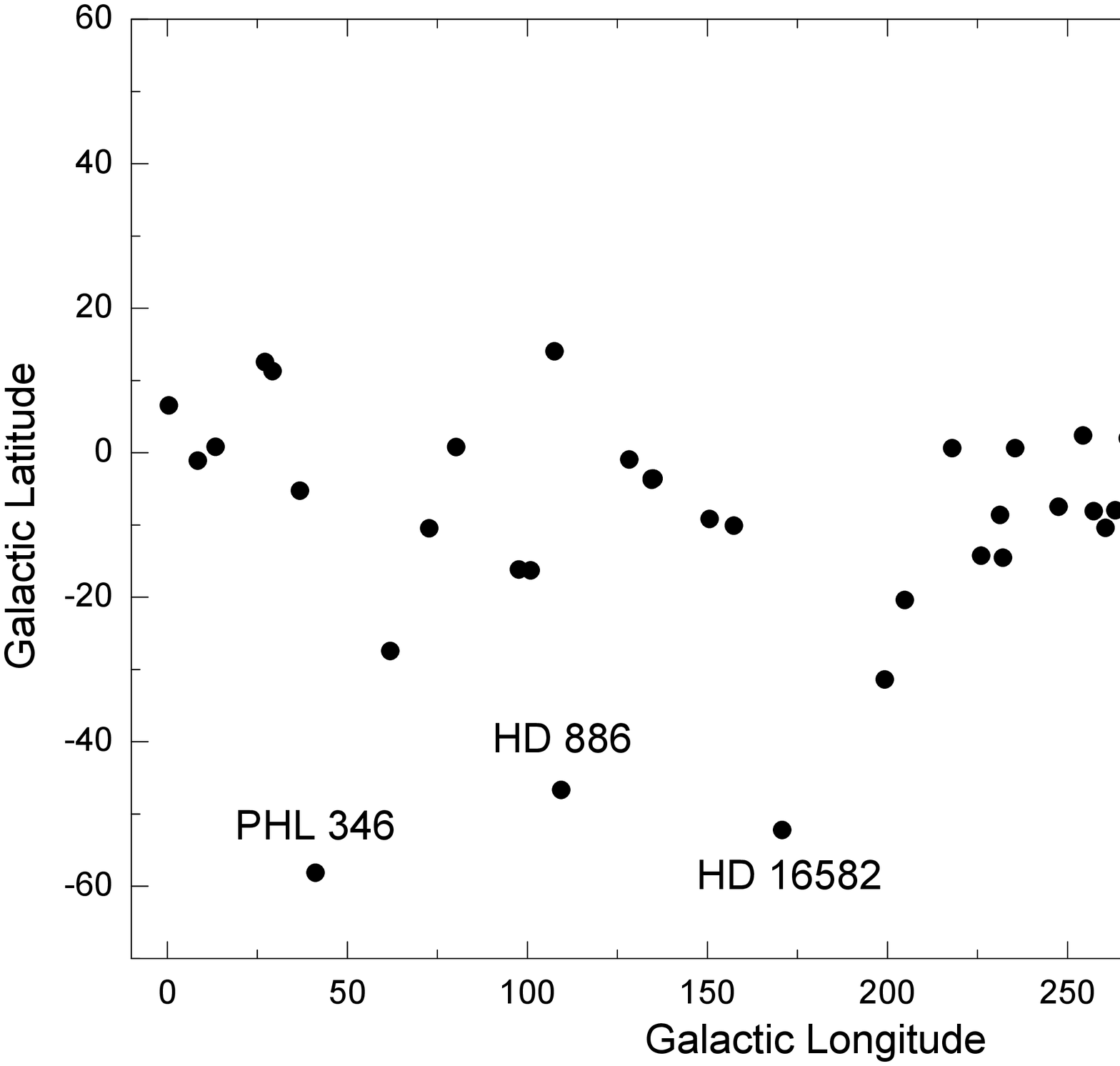}
\caption{Distribution of stars according to Galactic longitude and latitude.}
\label{gal}
\end{figure}

The Galactic distribution of the confirmed $\beta$~Cephei stars is shown in Fig.~\ref{gal}. In agreement with the result from Sect.\,5.1.3, this again suggests a young disk population. The most interesting objects in this diagram are the ``outliers,'' the only significant one being PHL~346, which may either have formed in the Galactic halo or could be a runaway star \citep{RHM01}.

\subsubsection{Pulsation amplitude vs. rotation rate and pulsation period vs. rotation rate}

\begin{figure}[h]
 \vspace{5.5cm}
 \includegraphics{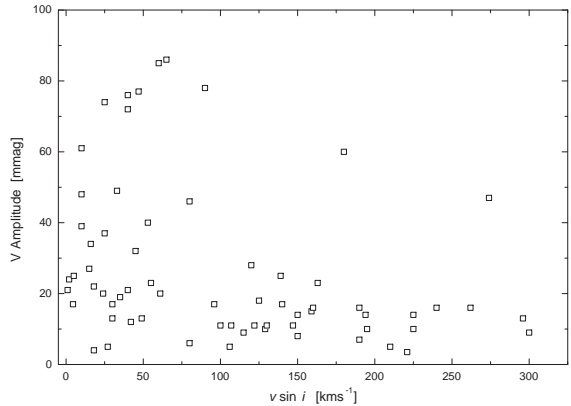}
\caption{Photometric amplitudes of the $\beta$~Cephei stars depending on their projected rotational velocity.}
\label{Avsini}
\end{figure}

The dependence between pulsation amplitude and rotation rate is plotted in
Fig.~\ref{Avsini}. With the exception of HD 52918 and HD 203664, only
stars with rotation velocities $v \sin i \leq 90$\,km\,s$^{-1}$ show
pulsation amplitudes larger than $\sim$ 25~mmag. This is similar to the
behavior of the $\delta$~Scuti stars \citep{Br82}, and may also lend
support to the hypothesis that rotation is an important factor in the
amplitude limiting mechanism operating in these types of pulsators. In
this context it is interesting to note that the range of the {\it angular}
rotational velocities of high-amplitude $\delta$~Scuti stars is very
similar to that of the $\beta$~Cephei stars with the highest amplitudes.
For a similar analysis based on the radial velocity pulsation amplitude we
refer to \cite{AD03}.

We also examined the pulsation period versus rotation velocities $v \sin
i$ and find that there is no dependence between these two quantities. This 
is also not a surprise as most of the known $\beta$~Cephei stars are 
photometric variables and thus pulsate in modes of low spherical degree.

\subsubsection{Mono- vs. multiperiodicity}

As listed in Table~4, $\sim$40\% of the confirmed $\beta$~Cephei stars are mono periodic. We suspect that several of these 37 stars may have additional pulsation periods that are undetected so far. On the other hand, our practical criteria to select $\beta$~Cephei stars are likely to introduce a bias in favor of multiperiodic stars. In any case, it seems safe to say that most $\beta$~Cephei stars are multiperiodic pulsators.

\subsubsection{Binarity}

Table 1 also contains information on binary $\beta$~Cephei stars, which
are indicated in the Notes column. Summarizing, we can say that there
are eight spectroscopic binaries, four additional double-lined spectroscopic
binaries, two suspected binaries, one eclipsing binary and one triple
system. Thus, we find that $\approx$14\% of all $\beta$~Cephei stars are 
located in known multiple systems with physically associated companions.

A search for visual binaries in The Catalogue of Components of Double and
Multiple Stars \citep{CCDM02} and the {\em Hipparcos} and Tycho Catalogs
\citep{ESA97} reveals that 16 stars of Table~1 are visual binaries. Five
of these are already known to be spectroscopic binaries as well. Owing to
these small numbers, we refrain from any statistical analysis. We also
assume that several additional $\beta$~Cephei stars will be proven to be
spectroscopic binaries in the future.

\subsection{HR-diagram, masses, pulsation constants and period-luminosity relation}

\begin{figure}[h]
 \vspace{7.5cm}
 \includegraphics{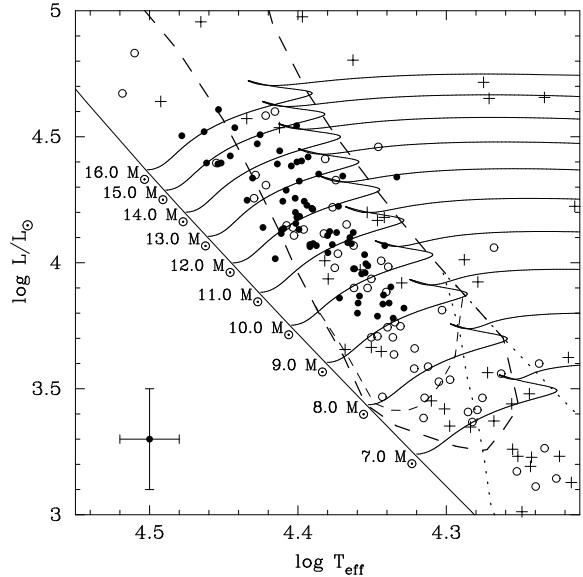}
\caption{Theoretical HR diagram of the confirmed ({\em filled circle}) and candidate ({\em open circles}) $\beta$~Cephei stars as well as the poor and rejected candidates ({\em plus signs}). The filled circles with the error bars in the lower left corner indicate the rms accuracy of each point in this diagram. The slanted solid line is the ZAMS, the thick dashed line describes the boundaries of the theoretical $\beta$~Cephei instability strip for $Z=0.02$, the thin dashed lines are the $\beta$~Cephei boundaries for radial modes, and the dotted lines those of the SPB stars. Several stellar evolutionary tracks, labeled with their evolutionary masses, are also plotted. All the theoretical results were adopted from \cite{P99}.}
\label{bcephrd}
\end{figure}

To obtain more insight into the behavior of the $\beta$~Cephei stars as a group, and for purposes of comparison with theoretical results, we have computed their temperatures and luminosities to place them in the HR diagram. To this end, we adopted the programs by \cite{NSW93} (which can be used for B stars of all luminosity classes), using published Str\"omgren indices from the GCPD to derive $T_{\rm eff}$ via the calibration by \cite{MD85} and $M_v$ from \cite{BS84}. We did not use {\em Hipparcos} parallaxes, as accurate results are only available for a few stars and as we wanted to treat the whole sample homogeneously. We then determined bolometric corrections from the work by \cite{F96}. The theoretical HR diagram constructed with these results is shown in Fig.~\ref{bcephrd}. The error estimates are $\pm$0.020 in $T_{\rm eff}$ and $\pm$0.20 in log\,L, which are hoped to include external uncertainties in the applied calibrations themselves.
 
We have also plotted the candidate $\beta$~Cephei stars (Table~2) and the
rejected candidates (Table~3) in this diagram for comparison. We compared
the positions of the stars in Fig.~\ref{bcephrd} with evolutionary tracks,
which we computed with the Warsaw-New Jersey stellar evolution code \cite[see, e.g., ][]{PDH98}. This way we estimated the masses of these
objects and we could consequently also compute the pulsation ``constant''
$Q$. The pulsation constant was derived from the period with the highest
amplitude value. Given the uncertainties in our determinations of $T_{\rm
eff}$ and $L$, we estimate an uncertainty of $\pm$ 30\% in $Q$. The errors
on $T_{\rm eff}$ and L should dominate the error introduced by not being
able to use the frequencies in the co-rotating frame for most stars. This
inability is due to missing mode identifications.

We adopted the theoretical boundaries of the \linebreak $\beta$~Cephei instability strip from the work by \cite{P99}. We prefer his results over those by \cite{DX01} because he applied newer versions of opacity tables and more reliable interpolation routines. The differences between these two approaches are discussed by \cite{P02} in detail.

The confirmed $\beta$~Cephei stars occupy a well-defined region in this plot with the exception of HD 165174 which appears to be so hot and luminous that it falls outside the boundaries of Fig.~\ref{bcephrd}. In contrast, the candidates and rejected stars are widely scattered. We note that the theoretically predicted instability strip is not completely filled with stars, a well-known problem that we will discuss in the next section.

In addition, a gap between the coolest $\beta$~Cephei stars at a given mass and the theoretical TAMS may be suspected. It is unclear whether this is a real feature or whether the derived absolute magnitudes from the Str\"omgren indices could be biased. \cite{HWS94} discussed this problem in detail. In any case, it is reasonable to conclude that all known $\beta$~Cephei stars are main-sequence objects. Consequently, the assignment of luminosity classes I--III to several confirmed \linebreak $\beta$~Cephei stars must be erroneous.

\begin{figure}[ht]
 \vspace{13cm}
 \includegraphics{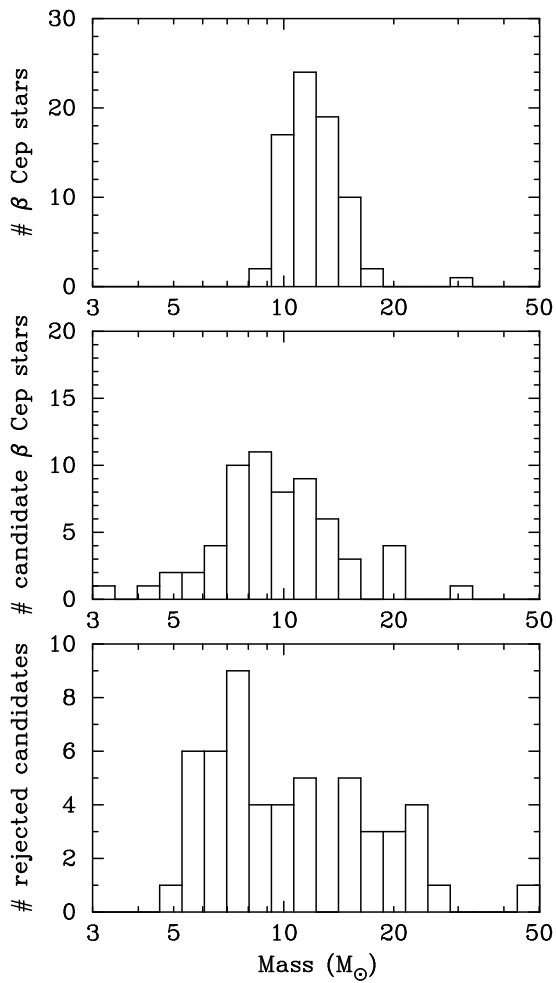}
\caption{Distribution of the masses of the stars in Tables 1--3.}
\label{M}
\end{figure}

We can now also examine the mass distribution of the $\beta$~Cephei stars and candidates (Fig.~\ref{M}). The mass of the confirmed $\beta$~Cephei stars peaks sharply at about 12 $M_{\sun}$. Whereas there is a slight indication for a similar maximum for the candidate $\beta$~Cephei stars, the histogram of the masses of the rejected stars is featureless.

\begin{figure}[ht!]
 \vspace{13cm}
 \includegraphics{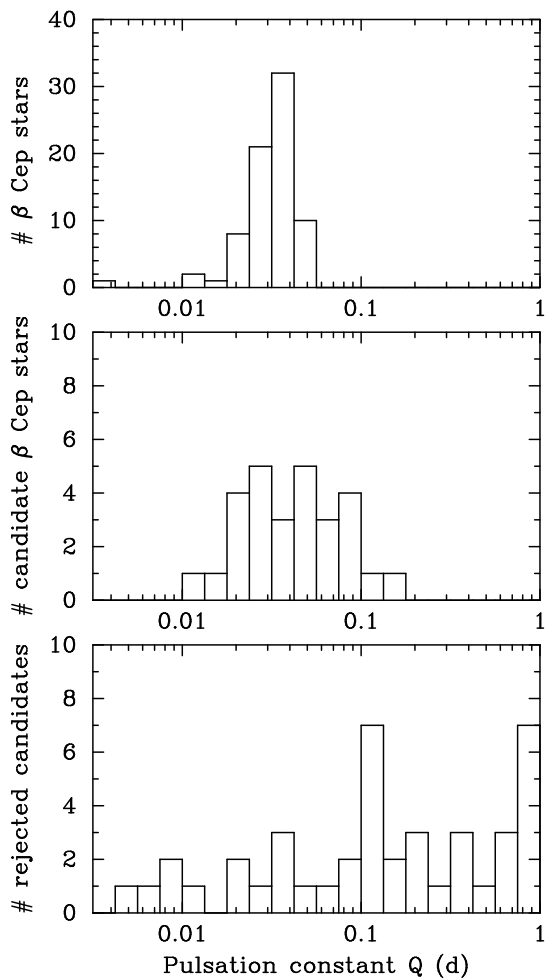}
\caption{Distribution of the pulsation constant $Q$ of the stars in Tables 1--3.}
\label{qhist}
\end{figure}

Turning to the pulsation constant (Fig.~\ref{qhist}), we again see a sharp peak for the confirmed \linebreak $\beta$~Cephei stars located at $Q$=0.033\,d, corresponding to the value for radial fundamental mode pulsation. More than half of the candidate $\beta$~Cephei stars have $Q$-values in the same range, although there is a tail toward higher $Q$. We remind that several stars were classified as candidate $\beta$~Cephei stars because of the lack of evidence that they are {\it not} pulsators. The histogram of $Q$ of the rejected candidates shows no particular preferences.  It is clear that $Q$-values for non pulsating stars have no real relevance, but our aim here is to check whether our separation of the candidates in the three groups was successful. Comparing the different panels within Fig.~\ref{M} and Fig.~\ref{qhist}, respectively, implies that the choice of our selection criteria is justified.

\section{The observational boundaries of $\beta$~Cephei pulsation}

As mentioned in Sect.~2 and 5.2, several authors computed the instability region for $\beta$~Cephei stars. Linear non-adiabatic analyses for low-degree \linebreak($\ell \leq$ 2) modes predict that photometrically observable modes are also driven in slightly evolved O-type stars \citep[e.g.,][]{DP93a}, suggesting that there could be a population of late O-type $\beta$~Cephei stars.

In 1998, the central region of the Cygnus OB2 association was investigated
in search of short-period hot pulsators \citep{PK98}. No $\beta$~Cephei
type stars were found among the O-type variables. So far, only one O-type
star, HD 34656 (O7e III) has been suggested to exhibit pulsations in the
$\beta$~Cephei domain \citep{FGB91}. Pulsation was claimed from radial
velocity measurements; the given period of 8.81\,hr is a little above the
typical range of pulsation periods for these stars. The authors were
reluctant to identify this star as a $\beta$~Cephei star. In addition, we
are unsure whether the reported radial velocity variations of the star are
statistically significant. Therefore, we cannot accept this O-star as a
confirmed $\beta$~Cephei star and place it therefore in Table~2.

There have been several similar attempts to discover O-type $\beta$~Cephei pulsators observationally \citep[e.g.,][]{B92}. However, to date no convincing detections were made, and, with the exception of the Be star HD~165174, there is consequently an apparently well-defined high-mass edge to the population in the resulting HR-diagram (Fig.~\ref{bcephrd}).

\begin{figure}[h]
 \vspace{10.5cm}
 \includegraphics{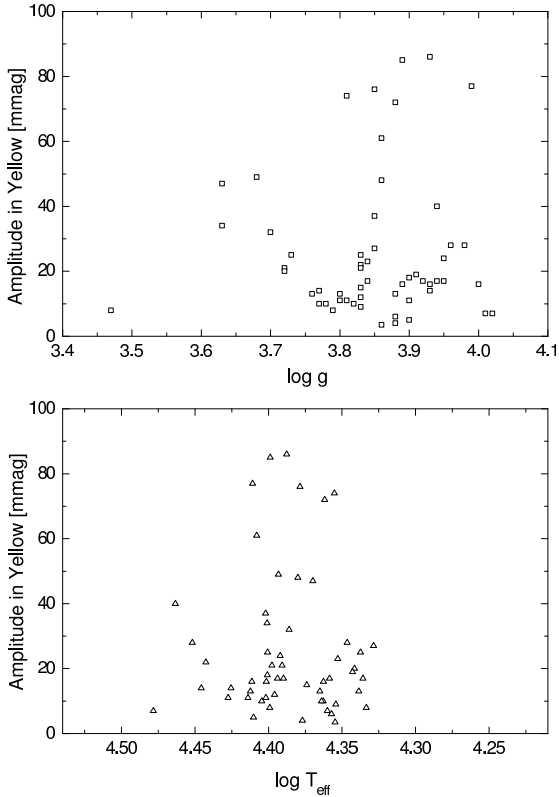}
\caption{{\em Upper panel}: log\,{\em g} vs. amplitude. {\em Lower panel}: log\,T$_{{\rm eff}}$ vs. amplitude. The {\em x}-axis here shows the same scale as the {\em x}-axis in Fig.~\ref{bcephrd}. The amplitudes in Johnson $V$ and Str\"omgren $y$ are shown for all stars where Str\"omgren indices were available. The amplitudes of the strongest modes are shown here as listed in Table~\ref{tbl-4}. }
\label{ATlogg}
\end{figure}

From Fig.~\ref{bcephrd} we see that the blue edge is a cut off for stars more luminous than log\,L$_{\odot}$=4.6 and hotter than log\,T$_{{\rm eff}}$=4.48. This result can be compared directly with Fig.~\ref{ATlogg} where we show the pulsation amplitudes versus log\,g and log\,T$_{{\rm eff}}$ ({\em upper and lower panel}, respectively). In the lower panel we see that the highest pulsation amplitudes occur in the middle of the instability region, as is expected because of the strong dependence of the $\kappa$-mechanism on temperature and hence on the depth of the ionization layer in which it operates. This diagram also suggests that O-type $\beta$~Cephei stars could exist, but that their pulsation amplitudes are small and therefore not yet detectable. Space missions could enable us to detect such pulsators.

There could be many reasons for the lack of observed O-type $\beta$~Cephei stars, as mentioned above. The theoretical models may not necessarily predict the real behavior of the stars, as some physics may be missing from the models. For example, the linear approach taken in the calculation of pulsation instability may not realistically reflect the complex physical processes in real stars, such as the onset of strong stellar winds. As mentioned before, it is also possible that O-type $\beta$~Cephei stars do exist, but with amplitudes below the current detection limits. In combination, these factors could prohibit the detection of O-type $\beta$~Cephei stars \citep[see also][]{PK98}.

In a recent publication, \cite{TMD03} analyze a sample of 49 presumable $\beta$~Cephei stars and show a HRD together with theoretical boundaries for the instability region computed by \cite{DX01}. Their computations also include a boundary at the high-mass end of the instability region, which stands in contrast to the theoretical work of \cite{P99} (see above). We compared their list of stars with our results and find that 27 of those stars are in our list of confirmed $\beta$~Cephei stars, 6 are classified as candidates and the remaining 16 are in the list of rejected stars. When we compare their HRD with Fig.~\ref{bcephrd}, we see that the boundaries adopted by \cite{TMD03} encompass all stars from Table~\ref{tbl-1}. Therefore, from an empirical point of view, both instability regions by \cite{P99} and \cite{DX01} fit our sample equally well.  

The theoretical $\beta$~Cephei instability strip is also not filled at the
low-mass (red) end. As the theoretical results seem to be more reliable in
this part of the HR diagram, the only explanation we have for this finding
is again that the pulsational amplitudes are too small to be detected by
current methods. We base this argument on analogy with the $\delta$~Scuti
stars, whose pulsations are of the same nature (low-order {\em p}- and {\em g}-modes
driven by the \linebreak $\kappa$-mechanism) as those of the $\beta$~Cephei stars, and whose number increases strongly with better detection levels \citep[see, e.g.,][Fig.~3]{Br79}. Support for this suggestion comes from intensive observations of individual $\beta$~Cephei stars \citep[see, e.g.][]{HSV03, J05}, for which more and more pulsation modes were
detected with decreasing detection threshold.

We note that the low-mass boundary of the theoretical instability strip in Fig.~\ref{bcephrd} for radial modes agrees better with the observations, but it is still too cool to be explained by errors in the temperature calibrations of the observed stars.
  
The empirical determination of the edges of the instability region is a
very interesting challenge in the field of $\beta$~Cephei stars. New
surveys of the late O-type stars and early to mid B-type stars would
therefore be of considerable astrophysical interest.

\section{Conclusions}

Of the 231 stars under consideration, 93 were confirmed as $\beta$~Cephei
type variable stars (Table~1). Their spectral types range from B0 to B3
with one exception, NGC~663\,4, whose spectral type of B5 does not appear
to be reliable. The periods of the strongest pulsation modes range from
$P$=0.0667\,d to 0.319\,d or 1.6\,h to 7.7\,h with a median of 0.171\,d.  
Projected rotational velocities $v\sin i$ range from 0 to 300\,km\,s$^{-1}$,
with a typical value of around 100\,km\,s$^{-1}$. This suggests that
$\beta$~Cephei stars are rather slow rotators, although this result could
be affected by a bias in detecting possible low-amplitude modes occurring
in rapidly rotating stars. We expect more detections of $\beta$~Cephei
stars concerning lower amplitude, higher $v\sin i$ stars with space
observations in the near future. The Galactic distribution of these stars
does not yield evidence for a pulsator that has formed at high Galactic
latitude.

There are 77 stars for which no clear classification could be made as a
result of limited or conflicting data. These are listed in Table~2 as
suspected $\beta$~Cephei type variable stars; they deserve further
attention. Additional notes are provided on interesting characteristics of
19 of these stars. Many of these stars seem good $\beta$~Cephei candidates
and it was often only the lack of recent data that forced us to put them
in the list of candidate $\beta$~Cephei stars.

Despite their previous classification as $\beta$~Cephei type variables or candidates, 61 of the stars could not be considered as such (Table~3). In some cases authors were over-confident in classifying them as $\beta$~Cephei stars. In other cases, later measurements have shown that they are either a different kind of variable, or that their variability is no longer detectable, casting doubt on the original observations. Some misclassifications are also due to historical reasons since, during the early days of work on $\beta$~Cephei stars, the group was not as well known as it is today.

The pulsation constant $Q$ calculated for all confirmed $\beta$~Cephei stars lies below Q=0.06\,d with a peak at Q=0.033\,d. The $Q$ value encompasses
many physical parameters, and its use as an observational constraint to
classify this group of variable stars is therefore considered more
accurate than previous classification techniques. These techniques often
relied more heavily on limited information such as spectral type
classification and pulsation period. This upper limit of Q=0.06\,d can
provide an additional observational constraint for the classification of
$\beta$~Cephei stars, keeping the uncertainties in the determination of $Q$ in mind.

The theoretical instability region for the \linebreak $\beta$~Cephei stars, as calculated by \cite{P99}, is not populated at both the low-mass/red end and the blue end. The lack of stars at the blue end, where one would expect late O-type stars, is expected \citep[e.g.,][]{B92, PK98}. We emphasize that this gap could be due to limitations in the theoretical modeling of the instability region, as well as to the difficulties inherent to observing hot stars exhibiting strong stellar winds and to possible pulsation amplitudes too low to be detected with past and present methods.

It is hoped that our new and refined catalog provides a useful framework within which to plan future observing campaigns, both ground-based and using the upcoming spaceborne observatories. The table of suspected $\beta$~Cephei variables provides a list of 77 interesting candidates that require further investigation. In addition, the catalog provides improved constraints on the classification and physical nature of $\beta$~Cephei variables, and these can in turn be used to correctly classify new early type short-period variable stars.

\section{Tables}

In Table~1 we present all confirmed $\beta$~Cephei stars. In Table~2 we list candidate $\beta$~Cephei stars and in Table~3 we give rejected $\beta$~Cephei candidates. At the end of each table we give notes on individual stars as well as short explanations on interesting characteristics of some stars. Table~4 contains a list of pulsation frequencies for all stars from Table~1. 

\subsection{Omitted stars}

Several candidate $\beta$~Cephei stars in Table~\ref{tbl-2} originate from the line profile variability surveys of \cite{TUI02} and \cite{STD02}. They were not directly claimed as $\beta$~Cephei candidates by these authors. Stars that show line profile variability but where we  discovered that the variations are likely not to originate in non-radial pulsation do not appear in the following tables at all. Those stars are: \\
{\bf \object{HD 11241} - 1 Per} - no periodicities in the spectroscopic data of \cite{JHL03}\\
{\bf \object{HD 48977} - 16 Mon} - probably a rotationally variable star\\
{\bf \object{HD 64503} - QZ Pup} - ellipsoidal variable with a residual variability of $P$$\sim$~1\,d, see \cite{HD86}\\
{\bf \object{HD 64740}} - rotational variable with a period of 1.33\,d, see \cite{L79} \\
{\bf \object{HD 154445}} - {\em Hipparcos} data analysis results in a period of 4.5916\,d with a peak-to-peak amplitude of 19\,mmag  \\
{\bf \object{HD 169467} - $\alpha$~Tel} - microvariable in {\em Hipparcos} with a period of 0.909\,d; it also is a He rich star and we suspect it to be a SPB star  \\
{\bf \object{HD 172910}} - {\em Hipparcos} data results in two periods: 1.1983\,d and 0.9812\,d and we suspect it to be a SPB star.

A similar comment applies to the $\zeta$ Ophiuchi stars listed in Table~1 of \cite{BD99}. Objects from that work which can have co rotating variability periods too long to be due to $\beta$~Cephei-type pulsation as described by us were not included in this catalog.

\clearpage
\begin{deluxetable}{cccccccccl}
\tabletypesize{\scriptsize}
\rotate
\tablecaption{Catalog of Galactic $\beta$~Cephei stars\label{tbl-1}}
\tablewidth{20cm}
\tablehead{
\colhead{HD} & \colhead{HIP} & \colhead{name} & \colhead{RA [h m s]} & \colhead{DE [$^o$ ' '']} & \colhead{Period [d]} & \colhead{V [mag]} & \colhead{MK} & \colhead{$v\sin i$ [km\,s$^{-1}$]} & \colhead{References}\\
\colhead{HR/Cluster} & \colhead{BD/CD} & \colhead{other name} & \colhead{(b-y)} & \colhead{m$_1$} & \colhead{c$_1$} & \colhead{$\beta$} & \colhead{RV [km\,s$^{-1}$]} & \colhead{Ampl. [mmag]} & \colhead{Notes}}
\startdata
 886 & 1067 & $\gamma$ Peg & 00 13 14 & +15 11 00 & 0.1518 & 2.8 & B2 IV & 4.5 & 005, 008, 051, 196 \\
 39 & +14 14 & Algenib & -0.107 & 0.093 & 0.116 & 2.627 & 3.5 & 17 $(V)$ & (10)\\
[3mm] & & & 01 36 39 & +61 25 54 & 0.207 & 10.5 & B1 III & & 145 \\
 & +60 282 & V909 Cas & & & & & & 50 $(R)$ & \\
[3mm] & & & 01 46 39 & +61 14 06 & 0.194 & 11.0 & B5 & & 187, 188\\
 NGC 663\,4 & & & 0.486 & -0.160 & 0.122 & 2.619 & & 40 $(V)$ & spectral type doubtful\\
[3mm] & & Oo 692 & 02 18 30 & +57 09 03 & 0.1717 & 9.3 & B0 V & & 019 \\
 NGC 869\,692 & +56 501 & V611 Per & & & & & & 19 $(V)$ & \\
[3mm] & & & 02 18 48 & +57 17 08 & 0.1949$\ast$ & 9.5 & B2 V & & 146 \\
 NGC 869\,839 & +56 508 & V665 Per & & & & & & 43 $(V)$ & \\
[3mm] & & Oo 992 & 02 19 00 & +57 08 44 & 0.1326 & 9.9 & B1 Vn & & 019 \\
 NGC 869\,992 & +56 520 & V614 Per & & & & & & 3 $(V)$ & \\
[3mm] & & Oo 2246 & 02 22 03 & +57 08 26 & 0.1842$\ast$ & 9.9 & B2 III & & 018 \\
 NGC 884\,2246 & +56 572 & & & & & & & 11 $(V)$ & \\
[3mm] & & Oo 2299 & 02 22 09 & +57 08 28 & 0.319 & 9.1 & B0.5 IV & 160 & 018, 147 \\
 NGC 884\,2299 & +56 575 & V595 Per & & & & & & 16 $(V)$ & (18) \\
[3mm] 16582 & 12387 & $\delta$ Cet & 02 39 28 & +00 19 42 & 0.1611 & 4.1 & B2 IV & 5 & 005, 008, 051, 084, 183 \\
 779 & -00 406 & 82 Cet & -0.099 & 0.091 & 0.102 & 2.616 & 12.7 & 25 $(V)$ & \\
[3mm] 21803 & 16516 & KP Per & 03 32 38 & +44 51 20 & 0.2018$\ast$ & 6.4 & B2 IV & 40 & 005, 008, 010, 242 \\
 1072 & +44 734 & & 0.082 & 0.023 & 0.102 & 2.617 & 2.4 & 72 $(V)$ & \\
[3mm] 24760 & 18532 & $\epsilon$ Per & 03 57 51 & +40 00 36 & 0.1603$\ast$ & 2.9 & B0.5 V & 130 & 051, 148, 149 \\
 1220 & +39 895 & 45 Per & -0.074 & 0.055 & -0.047 & 2.594 & 0.8 & 11 (Hp) &(10), (19) amplitude from this paper \\
[3mm] 29248 & 21444 & $\nu$ Eri & 04 36 19 & -03 21 08 & 0.1735$\ast$ & 3.9 & B2 III & 25 & 005, 051, 057, 062, 195, 230 \\
 1463 & -03 834 & 48 Eri & -0.076 & 0.068 & 0.072 & 2.610 & 14.9 & 74 $(y)$ & (10)\\
[3mm] 35411 & 25281 & $\eta$ Ori & 05 24 29 & -02 23 50 & 0.13 & 3.2 & B0.5 V & 130 & 051, 208, 236 \\
 1788 & -02 1235 & 28 Ori & -0.058 & 0.071 & -0.010 & 2.608 & 19.8 & n/a & (11), multiple system \\
[3mm] 35715 & 25473 & $\psi^2$ Ori & 05 26 50 & +03 05 44 & 0.0954$\ast$ & 4.6 & B2 IV & 141 & 051, 057, 177, 230\\
 1811 & +02 962 & 30 Ori & -0.088 & 0.075 & 0.033 & 2.619 & 12 & n/a & (3), (10), (20) \\
[6mm] 44743 & 30324 & $\beta$ CMa & 06 22 41 & -17 57 21 & 0.2513$\ast$ & 2.0 & B1 II-III & 1 & 008, 011, 051, 057, 079, 235 \\
 2294 & -17 1467 & Mirzam & -0.091 & 0.054 & -0.003 & 2.593 & 33.7 & 21 $(V)$ & (10) \\
[3mm] 46328 & 31125 & $\xi^1$ CMa & 06 31 51 & -23 25 06 & 0.2096 & 4.3 & B1 III & 16 & 005, 008, 010, 094 \\
 2387 & -23 3991 & 4 CMa & -0.093 & 0.064 & -0.022 & 2.585 & 26.9 & 34 $(V)$ & (10)\\
[3mm] 50707 & 33092 & EY CMa & 06 53 32 & -20 13 27 & 0.1846$\ast$ & 4.8 & B1 Ib & 49 & 005, 010, 056, 079, 084, 230 \\
 2571 & -20 1616 & 15 CMa & -0.087 & 0.071 & -0.014 & 2.594 & 28 & 13 $(V)$ & \\
[3mm] 52918 & 33971 & 19 Mon & 07 02 55 & -04 14 21 & 0.1912$\ast$ & 4.9 & B1 V & 274 & 005, 022, 051, 073, 174 \\
 2648 & -04 1788 & V637 Mon & -0.073 & 0.065 & 0.023 & 2.591 & 32 & 47 $(y)$ & (21) \\
[3mm] 56014 & 34981 & EW CMa & 07 14 15 & -26 21 09 & 0.0919 & 4.7 & B3 IIIe & 150 & 144, 174, 197 \\
 2745 & -26 4057 & 27 CMa & -0.067 & 0.070 & 0.168 & 2.572 & +16 & 8 $(V)$ & (10), (22)\\
[3mm] 59864 & 36500 & & 07 30 34 & -34 05 26 & 0.238$\ast$ & 7.6 & B1 III & & 009, 010, 094, 141 \\
 & -33 3879 & V350 Pup & 0.003 & 0.061 & 0.022 & 2.599 & 44 & 16 $(B)$ & \\
[3mm] 61068 & 37036 & PT Pup & 07 36 41 & -19 42 08 & 0.1664$\ast$ & 5.7 & B2 II & 10 & 005, 047, 057, 095, 149 \\
 2928 & -19 1967 & & -0.068 & 0.077 & 0.050 & 2.617 & 22 & 39 (b) & \\
[3mm] 64365 & 38370 & QU Pup & 07 51 40 & -42 53 17 & 0.1678$\ast$ & 6.0 & B2 IV & $\sim$30 & 005, 057, 094, 230 \\
 3078 & -42 3610 & & -0.075 & 0.076 & 0.112 & 2.622 & 32.2 & 13 $(V_W)$ & \\
[3mm] 64722 & 38438 & & 07 52 29 & -54 22 01 & 0.1034$\ast$ & 5.7 & B1.5 IV & 147 & 005, 094, 230 \\
 3088 & -54 1966 & V372 Car & -0.046 & 0.075 & 0.023 & 2.610 & 18 & 11 $(V_W)$ & \\
[3mm] 71913 & 41586 & YZ Pyx & 08 28 42 & -34 43 53 & 0.2058 & 7.7 & B1.5 II & & 058 \\
 & -34 4858 & & -0.012 & 0.052 & 0.024 & 2.594 & & 32 $(V_G)$ & \\
[3mm] 78616 & 44790 & KK Vel & 09 07 42 & -44 37 56 & 0.2157 & 6.8 & B2 II-III & 10 & 005, 009, 010, 051, 199 \\
 & -44 5150 & & 0.060 & 0.039 & 0.068 & 2.611 & 26 & 48 $(V)$ & (10)\\
[3mm] 80383 & & IL Vel & 09 17 31 & -52 50 19 & 0.1832$\ast$ & 9.2 & B2 III & 65 & 003, 005 \\
 & -52 2955 & & 0.097 & 0.013 & 0.072 & 2.617 & 19 & 86 $(V)$ & (10)\\
[3mm] 90288 & & & 10 23 57 & -57 27 52 & 0.1095$\ast$ & 8.2 & B2 III-IV & 240 & 003, 005, 094, 095 \\
 & -56 3324 & V433 Car & -0.040 & 0.054 & 0.020 & 2.622 & 4 & 16 $(V)$ & (10)\\
[3mm] 303068 & & & 10 34 48 & -58 08 54 & 0.1458$\ast$ & 9.8 & B1 V & 42 & 001, 005, 049 \\
 NGC 3293\,11 & -57 3329 & & 0.060 & 0.037 & 0.045 & 2.611 & -7 & 12 $(y)$ & \\
[6mm] 303067 & & & 10 35 30 & -58 12 00 & 0.1684$\ast$ & 9.5 & B1 V & 125 & 001, 005, 010, 049, 140 \\
 NGC 3293\,10 & -57 3340 & V401 Car & 0.082 & 0.034 & 0.047 & 2.604 & -10 & 18 $(y)$ & \\
[3mm] & & & 10 35 41 & -58 12 45 & 0.2506 & 8.7 & B1 IV & 33 & 001, 005, 010, 049, 140 \\
 NGC 3293\,16 & -57 3344 & V403 Car & 0.048 & 0.040 & 0.023 & 2.591 & -23 & 49 $(y)$ & \\
[3mm] & & & 10 35 45 & -58 14 00 & 0.1135 & 9.9 & B1 V & & 001, 140,143 \\
 NGC 3293\,65 & & V412 Car & 0.074 & 0.037 & 0.073 & 2.585 & & 8 $(y)$ & \\
[3mm] & & & 10 35 47 & -58 14 30 & 0.1621 & 9.2 & B1 III & 10 & 001, 094, 140 \\
 NGC 3293\,23 &  & V404 Car & 0.083 & 0.025 & 0.036 & 2.604 & 3 & 61 $(y)$ & \\
[3mm] & & & 10 35 48 & -58 12 33 & 0.1524$\ast$ & 9.3 & B0.5 V & 129 & 001, 005, 049, 140 \\
 NGC 3293\,14 & & V405 Car & 0.020 & 0.048 & 0.016 & 2.596 & -14 & 10 $(y)$ & \\
[3mm] & & & 10 35 54 & -58 14 48 & 0.16$\ast$ & 9.2 & B1 III & 194 & 001, 005, 009, 010, 049, 140 \\
 NGC 3293\,24 & & V378 Car & 0.089 & 0.025 & 0.006 & 2.593 & -12 & 14 $(y)$ & \\
[3mm] & & & 10 35 55 & -58 13 00 & 0.179 & 9.1 & B1 III & 225 & 001, 140 \\
 NGC 3293\,133 & & V440 Car & & & & & & 14 $(y)$ & \\
[3mm] & & & 10 35 58 & -58 12 30 & 0.1756$\ast$ & 9.3 & B1 V & 40 & 001, 005, 009, 010, 049, 140 \\
 NGC 3293\,18 & & V406 Car & 0.038 & 0.050 & 0.045 & 2.605 & -16 & 21 $(y)$ & (10)\\
[3mm]& & & 10 36 02 & -58 15 10 & 0.227 & 8.9 & B0.5 III & 61 & 001, 005, 009, 010, 140, 143\\
 NGC 3293\,27 & -57 3351 & V380 Car & 0.122 & -0.001 & 0.073 & 2.60 & -15 & 20 $(y)$ & \\
[3mm] 92024 & & & 10 36 08 & -58 13 05 & 0.1773$\ast$ & 9.0 & B1 III & 122 & 001, 005, 009, 049, 140, 215 \\
 NGC 3293\,5 & -57 3354 & V381 Car & 0.035 & 0.035 & 0.014 & 2.598 & -16 & 11 $(y)$ & (11) \\
[3mm] 109885 & 61751 & KZ Mus & 12 39 19 & -71 37 18 & 0.1706$\ast$ & 9.0 & B2 III & 47 & 003, 012, 050, 058 \\
 & -70 955 & & 0.173 & -0.010 & 0.060 & 2.620 & -61.1 & 77 $(V)$ & \\
[3mm] 111123 & 62434 & $\beta$ Cru & 12 47 43 & -59 41 19 & 0.1912$\ast$ & 1.3 & B0.5 IV & 18 & 008, 009, 051, 052, 084, 101 \\
 4853 & & Mimosa & -0.103 & 0.061 & -0.041 & 2.596 & 15.6 & 22 $(V)$ &(10) \\
[3mm] & & BS Cru & 12 53 21 & -60 23 21 & 0.1508$\ast$ & 9.8 & B0.5 V & 27 & 001, 004, 013 \\
 NGC 4755\,G (7) & -59 4454 & & 0.175 & -0.018 & 0.056 & 2.609 & -23 & 5 $(V)$ & \\
[3mm] & & & 12 53 26 & -60 22 26 & 0.2332 & 10.2 & B1 V & 106 & 001, 002, 004, 013 \\
 NGC 4755\,113 & & & 0.172 & 0.019 & 0.116 & 2.632 & -19 & 5 $(B)$ & \\
[6mm] & & & 12 53 38 & -60 22 39 & 0.1252$\ast$ & 10.2 & B2 V & 18 & 004, 013, 053 \\
 NGC 4755\,405 & & & 0.146 & 0.023 & 0.090 & 2.613 & -18 & 4 $(V)$ & \\
[3mm] & & CT Cru & 12 53 44 & -60 22 29 & 0.1305 & 9.8 & B1 V & 225 & 001, 002, 004, 013 \\
 NGC 4755\,301 & & & 0.179 & -0.021 & 0.103 & 2.605 & -6 & 10 $(V)$ & \\
[3mm] & & CV Cru & 12 53 47 & -60 18 47 & 0.1789$\ast$ & 9.9 & B1 Vn & 296 & 001, 004, 013 \\
 NGC 4755\,I (9) & & & 0.227 & -0.023 & 0.112 & 2.607 & -32 & 13 $(V)$ & \\
[3mm] & & CZ Cru & 12 53 52 & -60 21 52 & 0.1589$\ast$ & 10.1 & B1 V & 262 & 001, 002, 004  \\
 NGC 4755\,202 & & & 0.148 & 0.020 & 0.113 & 2.617 & -33 & 16 $(V)$ & \\
[3mm] & & CX Cru & 12 53 52 & -60 22 15 & 0.1825 & 9.4 & B1 V & 195 & 001, 004, 013 \\
 NGC 4755\,201 & & & 0.153 & 0.005 & 0.102 & 2.609 & -16 & 10 $(V)$ & \\
[3mm] & 62937 & CY Cru & 12 53 52 & -60 22 28 & 0.1592 & 9.7 & B1 V & 107 & 001, 002, 004, 013 \\
 NGC 4755\,307 & & & 0.174 & -0.028 & 0.152 & 2.620 & -27 & 11 $(B)$ & \\
[3mm] & & & 12 53 53 & -60 21 46 & 0.0933 & 10.3 & B2 Vn & & 001, 004, 069 \\
 NGC 4755\,210 & & & 0.177 & 0.022 & 0.136 & 2.634 & & 7 $(V)$ & \\
[3mm] & 62949 & BW Cru & 12 53 58 & -60 24 58 & 0.2049$\ast$ & 9.1 & B2 III & 96 & 001, 004, 013 \\
 NGC 4755\,F (6) & & ALS 2816 & 0.143 & 0.002 & 0.062 & 2.605 & -22 & 17 $(V)$ & \\
[3mm]112481 & 63250 & & 12 57 36 & -49 46 50 & 0.2596$\ast$ & 8.4 & B2 Ib & & 005, 009, 055, 093 \\
 & -49 7513 & V856 Cen & & & & 2.604 & -19 & 34 $(V_G)$ & \\
[3mm] 116658 & 65474 & $\alpha$ Vir & 13 25 11 & -11 09 40 & 0.2717 & 0.9 & B1 IV & 160 & 008, 009, 051, 052, 057, 094 \\
 5056 & -10 3672 & Spica & -0.114 & 0.080 & 0.018 & 2.605 & 1 & var. & (14) \\
[3mm] 118716 & 66657 & $\epsilon$ Cen & 13 39 53 & -53 27 59 & 0.1696$\ast$ & 2.3 & B1 V & 159 & 005, 009, 051, 052, 230, 232 \\
 5132 & -52 5743 & & -0.094 & 0.058 & 0.043 & 2.608 & 3 & 15 $(V_W)$ & (10)\\
[3mm] 122451 & 68702 & $\beta$ Cen & 14 03 49 & -60 22 22 & 0.1535$\ast$ & 0.6 & B1 II & 139 & 009, 043, 051, 059, 063, 230 \\
 5267 & -59 5054 & Agena & -0.092 & 0.045 & -0.004 & 2.594 & 5.9 & 25 $(V)$ & (4), (23)\\
[3mm] 126341 & 70574 & $\tau^1$ Lup & 14 26 08 & -45 13 17 & 0.1774 & 4.6 & B2 IV & 15 & 005, 008, 009, 052, 064 \\
 5395 & -44 9322 & 1 Lup & -0.047 & 0.064 & 0.132 & 2.621 & -21.5 & 27 $(V)$ & (10)\\
[3mm] 129056 & 71860 & $\alpha$ Lup & 14 41 55 & -47 23 17 & 0.2598$\ast$ & 2.3 & B1.5 III & 24 & 005, 008, 009, 051, 064, 230 \\
 5469 & -46 9501 & & -0.086 & 0.071 & 0.080 & 2.604 & 5.2 & 20 $(V_W)$ & (10)\\
[6mm] 129557 & 72121 & BU Cir & 14 45 10 & -55 36 05 & 0.1276$\ast$ & 6.1 & B2 IV & 30 & 005, 009, 064, 121 \\
 5488 & -55 5809 & & 0.036 & 0.027 & 0.058 & 2.617 & -6.4 & 17 $(y)$ & (10)\\
[3mm] 129929 & 72241 & & 14 46 25 & -37 13 19 & 0.1431$\ast$ & 8.1 & B3 V & 2& 005, 055, 065, 066, 094 \\
 & -36 9605 & V836 Cen & -0.059 & 0.058 & 0.038 & 2.618 & 66 & 24 $(V_G)$ & \\
[3mm] 136298 & 75141 & $\delta$ Lup & 15 21 22 & -40 38 51 & 0.1655 & 3.2 & B2 IV & 221 & 009, 068, 084, 200, 230 \\
 5695 & -40 9538 & & -0.101 & 0.075 & 0.076 & 2.616 & 0.2 & 3.5 $(V)$ & \\
[3mm] 144470 & 78933 & $\omega^1$ Sco & 16 06 48 & -20 40 09 & 0.0667 & 3.9 & B1 V & 89 & 051, 092, 173, 230 \\
 5993 & -20 4405 & 9 Sco & 0.037 & 0.043 & 0.010 & 2.618 & -2.6 & n/a & (1) \\
[3mm] 145794 & & & 16 15 26 & -52 55 15 & 0.1599$\ast$ & 8.7 & B2 II-III & & 005, 093, 094, 095 \\
 & -52 7312 & V349 Nor & & & & 2.615 & & 28 $(V_G)$ & (2) \\
[3mm] 147165 & 80112 & $\sigma$ Sco & 16 21 11 & -25 35 34 & 0.2468$\ast$ & 2.9 & B1 III & 53 & 005, 051, 096, 097, 098, 230 \\
 6084 & -25 11485 & 20 Sco & 0.168 & -0.032 & 0.003 & 2.604 & -1 & 40 $(V)$ & (2) \\
[3mm] 147985 & 80653 & & 16 26 57 & -43 47 57 & 0.1323$\ast$ & 7.9 & B1.5 II-III & 80 & 005, 051, 095, 102, 103 \\
 & -43 10792 & V348 Nor & & & & & & 46 $(V)$ & \\
[3mm] & & Braes 929 & 16 53 55 & -41 52 15 & 0.0671 & 9.6 & B1 V & 100 & 034, 036, 099, 100 \\
 NGC 6231\,253 & -41 11018p & V945 Sco & 0.206 & -0.020 & 0.031 & 2.596 & var. & 11 $(y)$ & (3) \\
[3mm] & & Braes 930 & 16 53 59 & -41 48 42 & 0.1193$\ast$ & 9.8 & B2 V & 80 & 015, 034, 036 \\
 NGC 6231\,282 & & V1032 Sco & 0.201 & -0.015 & 0.131 & 2.617 & & 6 $(y)$ & (3) \\
[3mm] & & Braes 932 & 16 54 02 & -41 51 12 & 0.0988$\ast$ & 10.3 & B2 IV-Vn & 140 & 034, 035, 036, 099, 100 \\
 NGC 6231\,261 & -41 11028 & V946 Sco & 0.228 & -0.046 & 0.144 & 2.626 & -32 & 17 $(y)$ & (4) \\
[3mm] 326330 & & Braes 672 & 16 54 18 & -41 51 36 & 0.0878$\ast$ & 9.6 & B1 Vn & 210 & 015, 034, 036 \\
NGC 6231\,238 & & V964 Sco & 0.198 & -0.004 & 0.015 & 2.615 & -30, var & 5 $(y)$ & \\
[3mm] & & Braes 948 & 16 54 35 & -41 53 39 & 0.1079$\ast$ & 9.8 & B1 V & 190 & 034, 036, 099, 100 \\
 NGC 6231\,110 & -41 11056 & V947 Sco & 0.237 & -0.011 & 0.006 & 2.612 & & 7 $(y)$ & (3) \\
[3mm] 326333 & & Braes 675 & 16 54 43 & -41 49 36 & 0.1012$\ast$ & 9.6 & B1 Vn & 150 & 034, 035, 036, 100, 101 \\
 NGC 6231\,150 & -41 11059 & V920 Sco & 0.215 & -0.013 & 0.026 & 2.606 & -47, var & 14 $(y)$ & (4) \\
[3mm] 156327B & 84655 & & 17 18 23 & -34 24 31 & 0.146$\ast$ & 9.4 & WC7 + B0III & & 185 \\
 & -34 11622 & V1035 Sco & & & & & & 35 $(V)$ & (11) \\
[6mm] 156662 & & & 17 21 06 & -45 58 56 & 0.1689$\ast$ & 7.8 & B2 III & 190 & 005, 095, 102 \\
 & -45 11411 & V831 Ara & 0.200 & -0.044 & 0.074 & 2.614 & & 16 $(V_G)$ & \\
[3mm] 157056 & 84970 & $\theta$ Oph & 17 22 01 & -24 59 58 & 0.1405$\ast$ & 3.3 & B2 IV & 35 & 005, 095, 104, 105, 106, 230 \\
 6453 & -24 13292 & 42 Oph & -0.092 & 0.089 & 0.104 & 2.624 & -5.6, var? & 19 $(y)$ & (10)\\
[3mm] 157485 & 85189 & & 17 24 35 & -26 55 29 & 0.2212$\ast$ & 9.1 & B1.5 Ib & & 058 \\
 & -26 12112 & V2371 Oph & & & & 2.623 & & 48 $(V_G)$ & \\
[3mm] 158926 & 85927 & $\lambda$ Sco & 17 33 37 & -37 06 14 & 0.2137$\ast$& 1.6 & B2 IV & 163 & 005, 051, 107, 108, 109, 230 \\
 6527 & -37 11673 & Shaula & -0.105 & 0.072 & 0.074 & 2.613 & 18.6, var & 23 $(V)$ & (5), (24) \\
[3mm] 160578 & 86670 & $\kappa$ Sco & 17 42 29 & -39 01 48 & 0.1998 &2.4 & B1.5 III & 115 & 008, 051, 107, 108, 110 \\
 6580 & -38 12137 & & -0.100 & 0.073 & 0.073 & 2.613 & 0.2,var & 9 $(V)$ & (2), (25) \\
[3mm] 163472 & 87812 & & 17 56 18 & +00 40 13 & 0.1399$\ast$ & 5.8 & B2 IV-V & 120 & 005, 024, 095, 111, 112, 113, 230 \\
 6684 & +00 3813 & V2052 Oph & 0.128 & 0.017 & 0.145 & 2.630 & -17.6 & 28 $(V)$ & (6) \\
[3mm] 164340 & 88352 & & 18 02 33 & -40 05 16& 0.1529 & 9.3 & B2 IV-V &  & 229, 241 \\
  & -40 12092&  &  &  &  & 2.584 &  & 25 $(V)$&  \\
[3mm] 165174 & 88522 & & 18 04 37 & +01 55 08 & 0.2907 & 6.2 & B0 IIIn & 300 & 114, 115, 116, 117, 182, 242 \\
 6747 & +01 3578 & V986 Oph & 0.075 & 0.000 & -0.119 & 2.567 & +17, var & 9 (b) & (2), (7), (26) \\
[3mm] 165812 & 88884 & & 18 08 45 & -22 09 38 & 0.1759$\ast$ & 7.9 & B1.5 II & & 058, 118 \\
 & -22 4581 & V4382 Sgr & 0.079 & 0.029 & -0.001 & 2.611 & -24 & 28 $(V_G)$ & (8) \\
[3mm] 166540 & 89164 & & 18 11 48 & -16 53 38 & 0.233$\ast$ & 8.1 & B1 Ib & 55 & 119 \\
 & -16 4747 & V4159 Sgr & & & & & -1.6 & 23 $(V)$ & \\
[3mm] 180642 & 94793 & & 19 17 15 & +01 03 34 & 0.1822 & 8.3 & B1.5 II-III & 90 & 058, 118, 242 \\
 & +00 4159 & V1449 Aql & 0.259 & -0.035 & 0.031 & & -14 & 78 $(V_G)$ & \\
[3mm] & & & 20 22 59 & +40 45 39 & 0.1565$\ast$ & 10.8 & B1 V & & 194  \\
 NGC 6910\,18 & & & 0.600 & -0.110 & 0.140 & 2.636 & &  15 $(V)$\\
[3mm] & & & 20 23 07 & +40 46 56 & 0.1922$\ast$ & 10.7 & B3 & & 194 \\
 NGC 6910\,16 & & & & & & & &  17 $(V)$\\
[3mm] & & & 20 23 08 & +40 46 09 & 0.1904 & 10.4 & B0.5 V & & 194 \\
 NGC 6910\,14 & & & 0.670 & -0.160 & 0.110 & 2.612 & &  17 $(V)$\\
 [6mm] & & & 20 23 34 & +40 45 20 & 0.143 & 11.8 & & & 194 \\
 NGC 6910\,27 & & & 0.820 & -0.180 & 0.170 & 2.625 & &  9 $(V)$\\
[3mm] & & V2187 Cyg & 20 33 18 & +41 17 39 & 0.2539 & 15.4 & & & 176\\
 & & & & & & & & 34 $(I)$ & \\
[3mm] 199140 & 103191 & BW Vul & 20 54 22 & +28 31 19 & 0.201 & 6.5 &B2 III & 60 & 005, 007, 051, 120, 121, 122 \\
 8007 & +27 3909 & & -0.033 & 0.051 & 0.029 & 2.610 & -8.5 & 85 $(V)$ & \\
[3mm] 203664 & 105614 & SY Equ & 21 23 29 & +09 55 55 & 0.1659$\ast$ & 8.6 & B0.5 IIIn & 180 & 058, 242 \\
 & +09 4793 & & & & & & 48 & 60 $(V_G)$ & \\
[3mm] 205021 & 106032 & $\beta$~Cep & 21 28 40 & +70 33 39 & 0.1905$\ast$ & 3.2 & B2 IIIe & 25 & 008, 051, 124, 125, 126, 127 \\
 8238 & +69 1173 & Alfirk & -0.092 & 0.066 & 0.010 & 2.605 & -3.1 & 37 $(V)$ & (2), (6), (7), (9) \\
[3mm] & & & 22 12 34 & +57 15 29 & 0.2029$\ast$ & 11.9 & B1.5 V & & 186\\
 NGC 7235\,8 & & & & & & & & 29 $(V)$ & \\
[3mm] & & HN Aqr & 22 37 38 & -18 39 51 & 0.1523 & 11.5 & B2 & 45 & 005, 095, 128, 129, 130 \\
 & & PHL 346 & -0.068 & 0.070 & 0.094 & & 63 & 32 $(V)$ & \\
[3mm] 214993 & 112031 & DD Lac & 22 41 29 & +40 13 21 & 0.1931$\ast$ & 5.3 & B2 III & 40 & 005, 051, 131, 132, 133, 134 \\
 8640 & +39 4912 & 12 Lac & -0.034 & 0.052 & 0.050 & 2.609 & -12.5 & 76 $(y)$ & (2), (9) \\
[3mm] 216916 & 113281 & EN Lac & 22 56 24 & +41 36 14 & 0.1692$\ast$ & 5.6 & B2 IV & 30 & 005, 051, 135, 137, 138, 240 \\
 8725 & +40 4949 & 16 Lac & -0.047 & 0.066 & 0.092 & 2.629 & -13 & var & (2), (10), (11) \\
\enddata
\end{deluxetable}

\clearpage

\subsection{Short notes to Tables 1--3}

\noindent(1) Only spectroscopic variability detected so far\\
(2) Spectroscopic binary\\
(3) Double-lined spectroscopic binary\\
(4) Suspected binary \citep{GM01}\\
(5) Spectroscopic triple system\\
(6) Magnetic star\\
(7) Mild Be star\\
(8) Periods from {\em Hipparcos} photomatry and Geneva data disagree\\
(9) Star is located in the overlap region of the BD and CD catalogs BD-22
4581 = CD-22 12607\\
(10) Visual binary\\
(11) Eclipsing binary\\
(12) Secondary component of a system with a Wolf-Rayet primary\\
(13) Possible $\zeta$ Ophiuchi star\\
(14) Ellipsoidal variable\\
(15) Double system with possible T Tauri component\\
(16) Claimed variability not confirmed\\
(17) No convincing variability in {\em Hipparcos} photometry\\

\subsection{Notes on individual $\beta$~Cephei stars}

{\bf V595 Per.--} (18) The period of its light variation is somewhat long and there seems to
be only one. The position of this star in the HR diagram of \cite{KP97}
leads to a pulsation constant of 0.039\,d. In $\beta$~Cephei models the
value of Q for the radial fundamental mode is between 0.034 and 0.041\,d
(if one only looks at modes excited in solar-metallicity models, the
upper boundary decreases to 0.036\,d). Assuming twice the photometric
period as the rotation period of a possible rotationally variable star, we
derive a rotational velocity of $\sim$~800\,km\,s$^{-1}$. This value is
higher than the break-up velocity and excludes the possibility of
rotational variability. Therefore V595 Per is confirmed to be a $\beta$~Cephei
star.\\
{\bf HD 24760 - $\epsilon$ Per.--} (19) Preliminary results by K. Uytterhoeven (private communication) on
this star show that several frequencies are probably excited in
$\epsilon$~Per and that harmonics are also present. More research on this
star is currently in progress. See also \cite{H99,GKJ99}.\\
{\bf HD 35715 - $\psi^2$ Ori.--} (20) Is also an ellipsoidal variable. Pulsation was not detected
photometrically but in line profiles.\\
{\bf HD 52918 - 19 Mon.--} (21) This is also a Be star (H$\alpha$ emission discovered by \cite{I75})  
with a relatively high pulsation amplitude that may be connected to shock
phenomena in the atmosphere. \cite{BJM02} find 3 frequencies, 2 of them
are due to $\beta$~Cephei-type pulsation.\\
{\bf HD 56014 - 27 (BW) CMa.--} (22) \cite{B95a} lists this star as a periodic Be star with a period of
$P=1.262$\,d. Short-period pulsations were, however, detected by
\cite{BKr94}, who report the re-detection of a period of $P$=0.0918\,d.
Next to HD 52918, this would be the second star to exhibit Be as well as
$\beta$~Cephei type variability.  It is also a close optical double system
and therefore it is possible that the $\beta$~Cephei variability does not
originate in the Be star. More research on this star is needed.\\
{\bf HD 122451 - $\beta$ Cen.--}
(23) Very eccentric double-lined spectroscopic binary with two 
$\beta$~Cephei components.\\
{\bf HD 158926 - $\lambda$ Sco.--} (24) This is a triple system with a variable dominant period of around 4.679410\,cd$^{-1}$. There are three additional significant frequencies which can, however, be attributed to either the primary or the tertiary component of this system \citep{UWL04,UTA04}.\\
{\bf HD 160578 - $\kappa$ Sco.--} (25) K. Uytterhoeven (private communication) confirms one pulsation mode
at 4.99922\,cd$^{-1}$, together with its first harmonic. All other
additional frequencies mentioned in the literature can be explained by 
means of a rotational modulation effect between a non-radial mode and the
rotation of the star in presence of spots on the stellar surface, but a
pure non-radial pulsation model cannot be excluded at the time being \citep{HUA04}.\\
{\bf HD 165174 - V986 Oph.--} (26) This is by far the hottest, most massive and most luminous
$\beta$~Cephei star; it also has one of the longest periods. The nature of
this mild Be star has been discussed in detail by \cite{CBM89}, and we
agree with these authors that there is no compelling reason not to
consider it a $\beta$~Cephei star. It satisfies our definition of this group 
of pulsating stars.

\clearpage
\begin{deluxetable}{cccccll}
\tabletypesize{\scriptsize}
\tablecaption{Candidate $\beta$~Cephei stars\label{tbl-2}}
\tablewidth{16cm}
\tablehead{
\colhead{HD} & \colhead{HIP} & \colhead{name} & \colhead{RA [h m s]} & \colhead{V [mag]} & \colhead{Reference}\\
\colhead{HR/Cluster} & \colhead{BD/CD} & \colhead{other name} & \colhead{DE ($^o$ ' '')} & \colhead{MK} & \colhead{Notes}}
\startdata
 & & & 01 46 40 & 12.4 & 188\\
 NGC 663\,114 & & & +61 09 52 & -----& \\
[3mm]13494 & & AG+56 243 & 02 13 37 & 9.3 & 039, 077 \\
& & V352 Per & +56 34 14 & B1 III & \\
[3mm]14053 & & & 02 18 23& 8.5 & 039 \\
NGC 869\,612  & +56 498& & +57 00 37& B1 II  & \\
[3mm]14250 & & AG+56 292 & 02 20 16 & 9.1 & 039 \\
& +56 545 & V359 Per & +57 05 55 & B1 IV & \\
[3mm]& & AG+57 301 & 02 22 50 & 9.6 & 039 \\
& +56 589 & V360 Per & +57 30 42 & B1 III & \\
[3mm] 21856 & 16518 & & 03 32 40 & 5.9 & 024, 231 \\
  1074 & +34 674 & & +35 27 42 & B1 V & not variable in 024, (13) \\
[3mm] & & & 04 07 43 & 9.3 & 039, 042 \\
 NGC 1502\,37 & & & +62 19 39 & B1.5 V & (27)\\
[3mm]& & & 04 07 44  & 9.6 & 042 \\
 NGC 150\,26 & +61 675& & +62 10 04 & B2 IV  & \\
[3mm] 25638 & 19272 & & 04 07 51  & 6.9 & 042 \\
NGC 1502\,1 & +61 676A & &  +62 19 48  &  B0 III  & \\
[3mm] 32990 & 23900 & & 05 08 06 & 5.5 & 231 \\
  1659 & +24 755 & 103 Tau & +24 15 55 & B2 V & (2) \\
[3mm]34656 & 24957 & & 05 20 43  & 6.8 & 006, 022, 025, 182\\
& +37 1146& & +37 26 19  & O7e III & (28) \\
[3mm] 35149 & 25142 & 23 Ori & 05 22 50 & 5.0 & 231, 230 \\
  1770 & +03 871 & & +03 32 40 & B1 V & (13)\\
[3mm] 36166 & 25751 & & 05 29 55 & 5.7 & 231 \\
  1833 & +01 1032 & & +01 47 21 & B2 V & (13)\\
[3mm]36512 & 25923 &  $\upsilon$ Ori  & 05 31 56  & 4.6 & 014, 088, 184 \\
1855 & -07 1106 & 36 Ori&  -07 18 05  & B0 V  & (29) \\
[3mm] 36695 & 26063 & VV Ori & 05 33 31 & 5.4 & 231, 230 \\
  1868 & -01 943 & & -01 09 22 & B1 V & (2), (13) \\
[3mm] 36819 & 26248 & & 05 35 27 & 5.4 & 231, 233 \\
  1875 & +23 954 & 121 Tau & +24 02 23 & B2.5 IV & (13) \\
[3mm] 37756 & 26736 & & 05 40 51 & 4.95 & 231 \\
  1952 & -01 1004 & & -01 07 44 & B2 IV-V & (2) \\
[3mm] 38622 & 27364 & & 05 47 43 & 5.27 & 231, 234 \\
  1993 & +13 979 & 133 Tau & +13 53 59 & B2 IV-V & (15) \\
[3mm] 39291 & 27658 & & 05 51 22 & 5.3 & 071, 153, 231 \\
  2031 & -07 1187 & 55 Ori & -07 31 05 & B2 IV-V & not variable in 153, (13)\\
[3mm] 40494 & 28199 & $\gamma$ Col & 05 57 32 & 4.4 & 230, 234 \\
  2106 & -35 2612 & & -35 16 59 & B2.5 IV & star in double system \\
[3mm]252248 & 29121 & AG+13 539 & 06 08 27 & 8.8 & 039, 077, 090 \\
NGC 2169\,5& & V917 Ori & +13 55 51 & B2 V & possible Be star \\
[3mm]43078 & 29687 & AG+22 667 & 06 15 15 & 8.8 & 039, 077 \\
& +22 1243 & LR Gem & 22 18 04 & B0 IV & (30) \\
[3mm] 44112 & 30073 & & 06 19 43 & 5.2 & 231, 230 \\
  2273 & -07 1373 & 7 Mon & -07 49 23 & B2.5 V & (2), (13) \\
[3mm] 45546 & 30772 & & 06 27 57 & 5.04 & 231 \\
  2344 & -04 1526 & 10 Mon & -04 45 44 & B2 V & star in double system \\
[3mm]51630 & 33447 & & 06 57 15 & 6.6 & 026, 043 \\
2603 & -22 1616 & & -22 12 10 & B2 III/IV & \\
[3mm] 53755 & 34234 & ADS 5782 A  &  07 05 50  &  6.5 &  039, 057, 087, 149 \\
 2670 & -10 1862 & V569 Mon &  -10 39 36  & B0.5 V & (31)\\
[3mm] 63949 & 38159 & QS Pup & 07 49 12 & 5.8  & 057, 095 \\
  3058 & -46 3460 & &  -46 51 27  & B1.5 lV & (32) \\
[3mm] 68324 & 39970 & IS Vel & 08 09 43 & 5.2  & 026, 057, 142 \\
  3213 & -47 3653 & & -47 56 13  & B2 IV & \\
[3mm] 69081 & 40321 & & 08 13 58 & 5.1 & 079, 230 \\
 3240 & -35 4358 & OS Pup & -36 19 20 & B1.5 IV & (13), in 079 slow variable \\
[3mm] 70839 & 40932 & & 08 21 12 & 5.9 & 230 \\
  3293 &  & & -57 58 24 & B1.5 III & (13)\\
[3mm] 70930 & 41039 & B Vel & 08 22 32 & 4.8 & 230 \\
  3294 & -48 3734 & & -48 29 25 & B1 V & double or multiple star, (13)\\
[3mm] 72108 & 41616 & & 08 29 05 & 5.3 & 230 \\
  3358 & -47 4004 & & -47 55 44 & B2 IV & double or multiple star \\
[3mm] 72127 & 41639 & & 08 29 28 & 4.99 & 230 \\
  3359 & -44 4462 & & -44 43 29 & B2 IV & double system, (13) \\
[3mm]74071 & 42459 & HW Vel & 08 39 24 & 5.4 & 043, 083 \\
3440 & & & -53 26 23 & B5 V & \\
[3mm] 74273 & 42614 & & 08 41 05 & 5.9 & 230 \\
  3453 & -48 4020 & & -48 55 22 & B1.5 V & (3), (13)\\
[3mm] 74455 & 42712 & HX Vel & 08 42 16 & 5.5 & 230, 237 \\
  3462 & -47 4251 & & -48 05 57 & B1.5 Vn & susp. ell. var in 237, (13), (33) \\
[3mm] 74575 & 42828 & $\alpha$ Pyx & 08 43 36 & 3.7 & 043, 178, 179\\
 3468 & -32 5651 &  & -33 11 11 & B1.5 III & (34), IR standard star\\
[3mm] 74753 & 42834 & D Vel & 08 43 40 & 5.1 & 230 \\
  3476 & -49 3761 & & -49 49 22 & B0 IIIn & (13)\\
[3mm] 86466 & 48799 & IV Vel  &  09 57 11 &  6.1 &  010, 079  \\
 3941 &   -52 3465 & &  -52 38 20  &   B3 IV  & (35)  \\
[3mm] 89688 & 50684 & RS Sex & 10 21 02 & 6.6 & 077, 078, 242 \\
  4064 & +03 2352 & 23 Sex & +02 17 23 & B3.2 IV & $P_{\it Hipparcos}\sim$0.129\,d\\ [3mm]96446 & 54266 & V430 Car & 11 06 06 & 6.7 & 075, 158 \\
& -59 3544 & & -59 56 59 & B2 I He & (36), He strong \\
[3mm] 97533 & 54753 & & 11 12 36 & 8.4 & 229 \\
   & -57 3772 & & -58 38 38 & B1:Vn & \\
[3mm]& &  & 11 21 09 & 12.4 & 045 \\
&&  V770 Cen & -60 32 13 & B5 e  & Be star\\
[3mm] & & &  11 36 14 & 9.9 & 165 \\
 NGC 3766\,67 & &  V847 Cen & -61 37 36 & B2Vp & (37)\\
[3mm] 104337 & 58587 & TY Crv & 12 00 51 & 5.3 & 023, 231 \\
  4590 & -18 3295 & 31 Crt & -19 39 32 & B1.5 V & (2), ell. var. in 023, (38) \\
[3mm] 108483 & 60823 & $\sigma$ Cen & 12 28 02 & 3.9 & 230, 153 \\
  4743 & -49 7115 & & -50 13 50 & B2 V & (13)\\
[3mm] 112092 & 63003 & $\mu$1 Cru & 12 54 36 & 3.9 & 230, 238 \\
  4898 &  & & -57 10 41 & B2 IV-V & double syst., not var. in 079 and 238 \\
[3mm] 120307 & 67464 &  $\nu$ Cen &  13 49 30  & 3.4 &  009, 051, 052, 076, 116, 198 \\
 5190 & & &  -41 41 15  & B2 V  & (39) \\
[3mm] 121743 & 68245 & $\phi$ Cen & 13 58 16 & 3.8 & 071, 079, 160, 230, 222 \\
  5248 & -41 8329 & & -42 06 03 & B2 IV & var. in 071, not var. in 079 and 160 \\
[3mm] 121790 & 68282 & $\upsilon$01 Cen & 13 58 41 & 3.8 & 160, 230, 222 \\
  5249 & -44 9010 & & -44 48 13 & B2 IV-V & (13), not var. in 160\\
 [3mm] 132058 & 73273 & $\beta$ Lup & 14 58 32 & 2.7 & 009, 051, 060\\
  5571 & -42 9853 & & -43 08 02 & B2 III & (13) \\
[3mm] 132200 & 73334 & $\kappa$ Cen & 14 59 10 & 3.1 & 230, 222 \\
  5576 & -41 9342 & & -42 06 15 & B2 IV & \\
[3mm] 136504 & 75264 & $\epsilon$ Lup & 15 22 41 & 3.4 & 230, 222 \\
  5708 & -44 10066 & & -44 41 23 & B2 IV-V & (2), (13)\\
[3mm] 142669 & 78104 & $\rho$ Sco & 15 56 53 & 3.9 & 076, 230 \\
  5928 & -28 11714 & 5 Sco & -29 12 51 & B2 IV-V & (2), (13)\\
[3mm]142883  & 78168 & & 15 57 40 & 5.9 & 017, 024, 076 \\
5934 & -20 4364& & -20 58 59 & B3V & \\
[3mm] 143018 & 78265 & $\pi$ Sco & 15 58 51 & 2.9 & 023, 076, 230 \\
  5944 & -25 11228 & 6 Sco & -26 06 51 & B1 V & (2), ecl. bin in 023, (40) \\
[3mm]144218 & 78821 & $\beta$ Sco A & 16 05 27 & 4.9 & 009, 029, 061, 149\\
5985 & -19 4308 & & -19 48 07 & B2 V & (41)\\
[3mm] 145502 & 79374 & $\nu$ Sco & 16 11 59 & 4.13 & 076, 230, 239\\
  6027 & -19 4333 & 14 Sco A & -19 27 39 & B2 IV & (2) \\
[3mm] 148703 & 80911 & & 16 31 23 & 4.22 & 230, 153, 222 \\
  6143 & -34 11044 & N Sco & -34 42 16 & B2 III-IV & \\
[3mm]149881 & 81362 &  & 16 36 58 & 7.0 & 027, 030, 031, 242 \\
& +14 3086 & V600 Her & +14 28 31 & B0.5 III & (42) \\
[3mm] 151985 & 82545 & $\mu$2 Sco & 16 52 20 & 3.5 & 230, 222 \\
  6252 & -37 11037 & & -38 01 03 & B2 IV & \\
[3mm]326327 & & Braes 669 & 16 53 39 & 9.7 & 034, 035 \\
NGC 6231\,28 &-41 11007 & V962 Sco & -41 47 48 & B1.5 Ve+sh & triple system? \\
[3mm]& & & 16 54 06 & 9.5 & 015, 036 \\
NGC 6231\,289& -41 11027p & & -41 51 13 & B0.5 V & \\
[3mm] & & & 16 54 14 & 10.3 & 015, 036 \\
 NGC 6231\,80& & V963 Sco & -41 55 01 & B0 Vn & \\
[3mm]& & & 16 54 16 & 10.2 & 036 \\
NGC 6231\,104 & & & -41 49 34 & -----& \\
[3mm]& & & 16 54 21 & 11.2 & 036\\
NGC 6231\,SBL\,515 & -41 7736& &-41 49 30 & B1 Vn & \\
[3mm] 163868 & 88123 &  & 17 59 56 & 7.4 & 229\\
  & -33 12700 &  &-33 24 29& B5 Ve & Be star\\
[3mm] 171034 & 91014 & & 18 33 58 & 5.3 & 230, 153, 222 \\
  6960 & -33 13338 & & -33 00 59 & B2 IV-V & (13)\\
[3mm] 176502 & 93177 & ADS 11910 A & 18 58 47 & 6.2 & 017 \\
  7179 & +40 3544 & V543 Lyr & +40 40 45 & B3 V & (43) \\
 [3mm]  & &  & 20 05 39 & 10.8 & 166 \\
 NGC 6871\,14 & & V1820 Cyg & +35 45 31 & B2 III & (44) \\
 [3mm]  & &  IC 4996 Hoag 7 & 20 16 45 & 10.9 & 024, 167, 168 \\
& &   V1922 Cyg & +37 40 44 & ----- & \\
[3mm] & & & 20 33 25 & 14.3 & 176 \\
& &  V2190 Cyg & +41 22 04 & -----& \\
[3mm]201819 & 104579 & & 21 11 04 & 6.5 & 024 \\
8105 & +35 4426 & & +36 17 58 & B0.5 IVn & \\
[3mm] 210808 & 109505 &  &  22 11 00 & 7.3 & 139 \\
&   +62 2045  &   V447~Cep &  +63 23 58  &  B5  & (45)\\
[3mm] & & & 22 54 17 & 15.9 & 189, 190\\
  NGC 7419\,BMD\,451 & & & 60 48 23 & -----& Be star \\
[3mm] & & & 22 54 19 & 16.4 & 189, 190\\
  NGC 7419\,BMD\,551 & & & 60 48 14 & -----& Be star \\
[3mm]217035 & & KZ~Cep & 22 56 31  & 7.7 & 039, 048 \\
& +62 2136 & & +62 52 07 & B0 V  & maybe Be star \\
\enddata
\end{deluxetable}

\newpage

\subsection{Notes on individual candidate $\beta$~Cephei stars}

{\bf NGC 1502\,37.--} (27) According to \cite{DAG92}, \cite{H67a} confused this star with NGC~1502\,A = NGC~1502\,1. We give its correct identification here and note in
addition that NGC 1502\,37 is a visual binary.\\
{\bf HD 34656.--} (28) This O7e III star was investigated by Fullerton et al. (1991) who detected
radial velocity variations with a period of 8.81\,h, of which we are
however not convinced. Fullerton et al. (1991) inferred that HD~34656 is a pulsating star and excluded the possibilities of the variations
originating in rotational modulation of a weak surface feature or motion
in a binary system. They associated its variability with $\beta$~Cephei
type pulsation but were reluctant to identify it as such a variable at
that time. This star is often cited to be the only O-type $\beta$~Cephei
star pulsator, despite the authors' caution.\\
{\bf HD 36512 - $\nu$ Ori.--} (29) Although the periods claimed for this star in the literature imply
SPB-like variability, our amplitude spectrum of its {\em Hipparcos} data has the
highest peak at a period of 0.146\,d indicating a $\beta$~Cephei nature of the pulsation.\\
{\bf HD 43078.-- } (30) \cite{H67a} suggests the presence of a fairly convincing 0.23887\,d
period for this star, which is however not present in the {\em Hipparcos} data. 
The Str\"omgren colors of this star are unusual, placing it
considerably below the ZAMS, and are inconsistent with its spectral
classification.\\
{\bf HD 53755 - V569 Mon.--} (31) \cite{B77} found a period of 0.18\,d. In the {\em Hipparcos} data we could
not detect any convincing periodicity. The highest peak in the amplitude
spectrum of these data is at 0.66\,d which is too long for $\beta$~Cephei
type pulsation.\\
{\bf HD 63949 - QS Pup.--} (32) There are doubts about the presence of the 0.1182\,d variation in the
1975 data set as well as about the 0.108\,d variation (C. Sterken,
private communication). The {\em Hipparcos} amplitude spectrum for this star
indicates no variability exceeding 4 mmag.\\
{\bf HD 74455.--} (33) \cite{M85} suspected it to be an ellipsoidal variable; confirmed in {\em Hipparcos} data (this work); see also \cite{WR83b}  \\
{\bf HD 74575 - $\alpha$ Pyx.--} (34) \cite{vH73Med} concluded from RV measurements that this star is a
$\beta$~Cephei variable; \cite{B77} found it not variable in RV whereas
\cite{SL83} found a well defined sinusoidal velocity curve with a probable
period of 5\,hr, but from one night only.\\
{\bf HD 86466 - IV Vel.--} (35) The available data are not conclusive. \cite{J79-552} places this
star in his "suspected $\beta$~Cephei stars" table. The highest peak in
the amplitude spectrum of the {\em Hipparcos} data of the star is at a 0.105\,d
period, but a 0.55\,d variation is almost equally probable.\\
{\bf HD 96446 - V430 Car.--} (36) This Bp star shows a 0.8514\,d period resulting from rotation, but a possible
secondary period near 0.26\,d could be due to pulsation \citep{MB91}.\\
{\bf NGC 3766\,67 - V847 Cen.--} (37) The frequency of the light variation of this candidate $\beta$~Cephei
star is close to 4 cycles per sidereal day, which could indicate an
extinction problem, and low-frequency variability also seems to exist.\\
{\bf HD 104337.--} (38) Ellipsoidal variability is confirmed by {\em Hipparcos} data (this work).  \\
{\bf HD 120307 - $\nu$ Cen.--} (39) This is a single lined spectroscopic binary and a Be star
 see Cuypers et al. (1989). The period of 0.4255\,d results in Q=0.107\,d, which is too large for $\beta$~Cephei pulsation. Most of the other periods found for this star are too long for $\beta$~Cephei pulsation as well.  \cite{ST02} however detected seven frequencies spectroscopically, that
they attributed to high degree modes ($\ell >$5) which could be connected
to $\beta$~Cephei type pulsation or be $\zeta$~Ophiuchi-like line profile
variability. \\
{\bf HD 143018.--} (40) Ellipsoidal variable with $P$=1.570\,d, see \cite{SLK96}.  \\
{\bf HD 144218 - $\beta$ Sco A.--} (41) Binary system; $\beta$~Cep candidate with a tentative period of
$P$=0.1733\,d \citep[see][]{HHH97}.\\
{\bf HD 149881 - V600 Her.--} (42) Possibly an ellipsoidal variable with a $\beta$~Cephei component \citep{DDU04}.  
Pulsational variability not detectable in {\em Hipparcos} data within a limit of 4.5 mmag.\\
{\bf HD 176502 - V543 Lyr.--} (43) Visual double star. The {\em Hipparcos} data clearly indicate that the star
is variable, but the time scale remains unknown due to aliasing; it could
be either several days or 2.5\,hr.\\
{\bf NGC 6871\,14 - V1820 Cyg.--} (44) Few variability measurements of this candidate $\beta$~Cephei star are 
available, and the star is under-luminous for the rather long period 
claimed.\\
{\bf HD 210808 - V447 Cep.--} (45) The analysis of this star's {\em Hipparcos} photometry reveals a primary
period of 0.314\,d, and a possible secondary period of 0.460\,d
\citep{K01}. The late spectral type of the star is inconsistent with its
Str\"omgren H$\beta$ index (2.639), suggesting a possible Be nature. The
star is also known as a visual binary and as an X-ray source.\\

\clearpage
\begin{deluxetable}{cccccll}
\tabletypesize{\scriptsize}
\tablecaption{Poor and rejected $\beta$~Cephei candidates\label{tbl-3}}
\tablewidth{16cm}
\tablehead{
 \colhead{HD} & \colhead{HIP} & \colhead{name} & \colhead{RA} & \colhead{V [mag]} & \colhead{Ref}\\
 \colhead{HR/Cluster} & \colhead{BD/CD} & \colhead{other name} & \colhead{DE ($^o$ ' '')} & \colhead{MK} & \colhead{Notes}}
\startdata
 3379 & 2903 & 53 Psc & 00 36 47 & 5.9 & 070, 071 \\
  155 & +14 76 & AG Psc & 15 13 54 & B2.5 IV & SPB star \\
[3mm]  & & & 01 32 37 & 9.5 & 228\\
 &  +61 285 & & +61 58 12 & B0.5 III: & $P \sim $2\,d, aliasing mistake in reference \\
[3mm] 13051 & 10541 & V351 Per & 02 09 26 & 8.6 & 072, 077\\
 &  +55 554 & & +56 59 30 & B1 IV: & $P_{\it Hipparcos}\sim$2.5\,d \\
[3mm] 13544 & 10391 & AG+53 218 & 02 13 52 & 8.9 & 017, 039, 084 \\
 +53 480 & & V353 Per & +53 54 53 & B0.5 IV & (46) \\
[3mm]& & & 02 16 58 & 9.2 & 028, 039, 072 \\
NGC 869\,49 & +56 473 & V356 Per& +57 07 49 & B0.5 IIIn & Be star, (16)\\
[3mm]  & & NSV 776 & 02 18 58  & 11.0 & 019 \\
 NGC 869\,963  & & Oo 963 & +57 08 18  &  B2 IV  & (16)\\
[3mm] 13745 & & AG+55 231 & 02 15 46 & 7.9 & 039, 242 \\
 & &  V354 Per & +55 59 47 & O9.7 II & (17)\\
[3mm] 13831 & 10615 & & 02 16 39 & 8.3 & 164 \\
 &  +56 469 & V473 Per & +56 44 16 & B0 IIIp & (47) \\
 [3mm] 13866 & 10641 & V357 Per & 02 16 58 & 7.5 & 039, 072 \\
 &  +56 475 & & +56 43 08 & B2 Ib & (16) \\
[3mm] 15239 & 11604&  & 02 29 38 & 8.5 & 040 \\
 St7-28 & +60 487 & V528 Cas& +60 39 26 & B2.5V+sh & $P_{\it Hipparcos}\sim$1\,d \\
[3mm]15752 & 11953 & AG+58 273 & 02 34 12 & 8.8 & 039 \\
& +57 589 & V362 Per & +58 24 20 & B0 III & (17)\\
 [3mm] 16429A & 12495 & STF 284A & 02 40 45 & 7.9 & 039, 152 \\
 &  +60 541 & V482 Cas & +61 16 56 & O9.5 III & (48) \\
 [3mm] 19374 & 14514 & 53 Ari & 03 07 26 & 6.1 & 078, 153, 243 \\
  938 & +17 493 & UW Ari & 17 52 48 & B1.5 V & (16) \\
[3mm]23480 & 17608 & Merope & 03 46 20 & 4.2 &  020, 021, 144 \\
 1156 & +23 522 & V971 Tau& +23 56 54 & B6 IVe  & periodic Be star, $P=0.49$\,d\\
[3mm]24640 & 18434 & NSV 1418  & 03 56 29  & 5.5 & 041, 231 \\
1215 & +34 768 & & +35 04 51  & B1.5 V  & (49) \\
[3mm] 27396 & 20354 & 53 Per & 04 21 33 & 4.8 & 154, 155 \\
  1350 & +46 872 & V469 Per & +46 29 56 & B4 IV & SPB star\\
[3mm]28114 & 20715 &  & 04 26 21 & 6.1 & 156 \\
1397 & +08 687 & V1143 Tau& +08 35 25 & B6 IV &SPB star\\
[3mm]28446 & 21148 &  DL Cam  & 04 32 01  & 5.8 & 023, 024 \\
1417 & +53 779 & 1 Cam & +53 54 39  & B0 IIIn & (50) probably SPB\\
[3mm] 33328 & 23972 & $\lambda$ Eri & 05 09 09 & 4.2 & 006, 084, 144, 182 \\
 1679 & -08 1040 & 69 Eri & -08 45 15 & B2 IVne & periodic Be star, $P=0.702$\,d \\
[3mm] 35468 & 25336 & $\gamma$ Ori & 05 25 08 & 1.6 & 180, 181\\
  1790 & +06 919 & 24 Ori & +06 20 59 & B2 III & (51)\\
[3mm] 37776 & 26742& & 05 40 56 & 6.9& 016, 039, 067 \\
&  -01 1005& V901 Ori & -01 30 26& B2 IV  & (52)\\
[3mm] 38010 & 26998 & & 05 43 39 & 6.8 & 157 \\
 &  +25 941 & V1165 Tau & 25 26 22 & B1 Vpe & Be star, $P_{\it Hipparcos}\sim$0.67\,d \\
[3mm] 252214 & 29106 & AG+13 535 & 06 08 18 & 8.1 & 039, 077, 090 \\
  NGC 2169\,2 & +13 1120 & V916 Ori & 13 58 18 & B2.5 V & (16) \\
[3mm]43837 & 30041 & AG+20 661 & 06 19 17 & 8.5 & 017, 074, \\
& +20 1369 & & +20 34 48 & B2 Ibp... & $P_{\it Hipparcos}\sim $2\,d \\
[3mm]43818 & 30046 & LU Gem & 06 19 19 & 6.9 & 074, 079, 089 \\
& +23 1300 & 11 Gem & +23 28 10 & B0 II & (53)\\
[3mm]47432 & 31766 & & 06 38 38  & 6.2 & 017, 043, 044 \\
2442 & +01 1443 & V689 Mon & +01 36 49  & O 9.5 III & $P_{\it Hipparcos}\sim $2\,d \\
[3mm]51309 & 33347 & $\iota$ CMa & 06 56 08 & 4.4 & 046, 071, 088 \\
2596 & -16 1661 & 20 CMa & -17 03 15 & B3 Ib/II & (54)\\
[3mm]53974 & 34301 & FN CMa & 07 06 41 & 5.4 & 039, 043, 077 \\
2678 & -11 1790 & & -11 17 39 & B0.5 IV & $P_{\it Hipparcos}\sim $1\,d\\
[3mm] 55857 & 34924 & GY CMa & +07 13 36 & 6.1 & 079, 191, 192 \\
  2734 & -27 3789 & ALS 255 & -27 21 23 & B0.5 V & (16) \\
[3mm] 55958 & 34937 & GG CMa & 07 13 47  & 6.6 & 026, 027 \\
 2741 & -30 4143 & & -03 01 51  & B2 IV  & (16) \\
[3mm] 57219 & 35406 & $\upsilon$02 Pup & 07 18 39 & 5.1 & 022, 086, 172 \\
  2790 & -36 3519 & NW Pup & -36 44 34 & B2 IVne & (55)\\
[3mm] 65575 & 38827 & $\chi$ Car & 07 56 47 & 3.4 & 046, 084, 085, 153 \\
  3117 & & & -52 58 57 & B3 IVp & (56) \\
[3mm] 67536 & 39530 & & 08 04 43 & 6.2 & 079, 081, 084 \\
  3186 & -62 330 & V375 Car & -62 50 11 & B2.5 Vn & Be star, $P_{\it Hipparcos}=1.01646$\,d \\
[3mm]74195 & 42536 & o Vel & 08 40 18 & 3.6 & 029 \\
 3447 & & & -52 55 19& B3 IV & SPB star \\
[3mm]74375 & 42568 & d Car & 08 40 37 & 4.3 & 017, 081, 082 \\
3457 & -59 2020 & V343 Car& -59 45 40 & B1.5 III & $P_{\it Hipparcos}=2.37952$\,d \\
[3mm] 74280 & 42799 & $\eta$ Hya & 08 43 14 & 4.9 & 046, 149, 150 \\
  3454 & +03 2039 & 7 Hya & +03 23 55 & B3 V & $P_{\it Hipparcos}\sim 2.2$\,d \\
[3mm] 77002 & 43937 & b01 Car & 08 56 58 & 4.9 & 079, 080, 160 \\
  3582 & -58 2347 & V376 Car & -59 13 45 & B2 IV-V & (16) \\
[3mm]77320 & 44213 & IU Vel & 09 00 22& 6.1 &  021, 022, 144  \\
 3593 & -42 4875& & -43 10 26& B2.5 Vne  & periodic Be star ($P$=0.612\,d)\\
[3mm]85953 & 48527& & 09 53 50& 5.9 & 029 \\
3924 & -50 4622& V335 Vel& -51 08 48 & B2 III & SPB star\\
[3mm] 92007 & & & 10 35 59 & 8.2 & 001, 005, 009, 049, 144 \\
  NGC 3293\,26 & -57 3350 & V379 Car & -58 14 15 & B1 III & periodic Be star ($P$=1.754\,d) \\
[3mm] 98410 & 55207 & ALS 2299 & 11 18 18 & 8.8 & 023 \\
 &  -62 505 & V536 Car & -62 58 28 & B2/B3 Ib/II & $P_{\it Hipparcos}=1.45325$\,d\\
 [3mm] 104841 & 58867 & $\theta^2$ Cru & 12 04 19 & 4.7 & 046, 079 \\
  4603 & -62 610 & & -63 09 57 & B2 IV & (57) \\
[3mm] 106490 & 59747 & $\delta$ Cru & 12 15 09 & 2.8 & 043, 046, 047, 151, 153, 230 \\
  4656 & -58 4466 & & -58 44 56 & B2 IV & (58)\\
[3mm] 109668 & 61585 & $\alpha$ Mus & 12 37 11 & 2.7 & 043, 046, 079, 153 \\
  4798 & -68 1104 & & -69 08 08 & B2 IV-V & (59)\\
[3mm] & & BT Cru  & 12 53 36 & 9.6 & 002, 032, 033, 193, 230\\
 NGC 4755\,418 & -59 4542 & & -60 23 46 & B2 V & (16)\\
[3mm]& & & 12 53 38 & 11.6 & 001, 004, 054\\
NGC 4755\,215 & & & -60 22 49 & & $P$=0.355\,d, SPB star?\\
[3mm]  & & BV Cru  & 12 53 39 & 8.7 & 002, 033, 193\\
 NGC 4755\,105 & & & -60 21 12 & B0.5 IIIn  &  $P\sim1$\,d or 2\,d, possible binary \\
[3mm] 112078 & 63007 & $\lambda$ Cru & 12 54 39 & 4.6 & 010, 022, 047, 077, 144 \\
  4897 &-58 4794 & & -59 08 48 & B4 Vne & $P_{\it Hipparcos}=0.35168$\,d, $Q=0.11$\,d \\
[3mm] 116072 &  & & 13 22 36 & 6.2 & 084, 159 \\
  5034 & -60 4639 & V790 Cen & -60 58 19 & B2.5 Vn & $\beta$ Lyr-type eclipsing binary\\
 [3mm]  122980 & 68862 & $\chi$ Cen & 14 06 03 & 4.4 & 160 \\
  5285 & -40 8405 & & -41 10 47 & B2 V & (60)\\
[3mm] 130903 & 72710 & He 3-1034 & 14 51 58 & 7.9 & 023, 053, 161 \\
 &  -40 9037 & V1018 Cen & -40 48 21 & B2p & (61)\\
[3mm] 160762 & 86414 & $\iota$ Her & 17 39 28 & 3.8 & 162, 163 \\
  6588 & +46 2349 & 85 Her & +46 00 23 & B3 IV & (62)\\
[3mm] 160124 & 86432 & & 17 39 38 & 7.2 & 014, 175 \\
  NGC 6405\,100 & -32 13072 & V994 Sco & -32 19 13 & B3 IV & SPB star\\
 [3mm] 180125 & 94588 & & 19 14 58 & 7.4 & 023, 161 \\
 &  +10 3839 & V1447 Aql & +10 24 34 & B8 V & $P_{\it Hipparcos}=2.1678$\,d \\
 [3mm] 180968 & 94827 & ES Vul & 19 17 44 & 5.4 & 084, 114, 144 \\
  7318 & +22 3648 & 2 Vul & +23 01 32 & B0.5 IV & periodic Be star ($P$=1.27\,d) \\
 [3mm] 188439 & 97845 & & 19 53 01 & 6.3 & 017, 114 \\
  7600 & +47 2945 & V819 Cyg & +47 48 28 & B0.5 IIIn & (63)\\
[3mm]189687 & 98425& 25 Cyg & 19 59 55 & 5.1 & 037, 038 \\
7647 & +36 3806 & V1746 Cyg & +37 02 34 & B3 IVe & Be star \\
[3mm] 195556 & 101138 & $\omega^1$ Cyg & 20 30 04 & 4.9 & 024 \\
  7844 & +48 3142 & 45 Cyg & +48 57 06 & B2.5 IV & (64) \\
[3mm]204076 &  105912 &BR Mic&  21 27 01  & 8.8 &  123 \\
&   -32 16569&& -31 56 20&  B2 II  & $P_{\it Hipparcos}\sim3.6$\,d\\
[3mm] 217811 & 113802 & LN And & 23 02 45 & 6.4 & 170 \\
  8768 & +43 4378 & & +44 03 32 & B2 V & (65) \\
[3mm] 224559 & 118214 &LQ And & 23 58 46 & 6.5 & 144, 169, 171  \\
  9070 & +45 4381 & AG+46 2225 & 46 24 47 & B4 Vne & periodic Be star ($P$=0.619\,d)
\enddata
\end{deluxetable}

\clearpage
\subsection{Notes on individual rejected $\beta$~Cephei stars}

{\bf HD 13544 - V353 Per.--} (46) Two periods of 0.6647 and 0.7724\,d explain this star's {\em Hipparcos}
photometry.\\
{\bf HD 13831 - V473 Per.--} (47) Be star. Published data indicate short-period variability, but 
{\em Hipparcos} photometry (this work) fails to confirm that.\\
{\bf HD 16429A - V482 Cas.--} (48) This star is a speckle binary in a triple system; also a radio
emitter and an X-ray source. Time scales present in its {\em Hipparcos} light 
curves are of the order of \linebreak $P$=1.7--2.5\,d.\\
{\bf HD 24640.--} (49) The published radial velocity curves are not convincing. Variability
time scales in the star's {\em Hipparcos} photometry are longer than 1.5\,d.\\
{\bf HD 28446 - 1 (DL) Cam.--} (50) Our analysis of this star's {\em Hipparcos} photometry results in candidate
periods considerably longer than those of $\beta$~Cephei type pulsation; 
it is also possible that parts of eclipses were observed by the satellite.\\
{\bf HD 35468 - $\gamma$ Ori.--} (51) \cite{K94} and \cite{KL96} suspect this is a low amplitude, possibly 
irregular variable. However, their measurements are too scarcely
sampled to enable a search for periods in the range of $\beta$~Cephei
pulsations.\\
{\bf HD 37776 - V901 Ori.--} (52) This is a rapidly rotating magnetic CP star \citep{CR98}. We
determine a period of 1.538\,d from its {\em Hipparcos} photometry.\\
{\bf HD 43818 - 11 (LU) Gem.--} (53) Most recent data \citep{P84} show no evidence for variability on a
time scale $<$0.2\,d but on a time scale of $>$0.2\,d or more likely
$>$0.5\,d. The period derived in that paper is $P$=1.25\,d. Period from
{\em Hipparcos} $\sim$2.1\,d (this work).\\
{\bf HD 51309 - $\iota$ CMa.--} (54) There are no new data since the work of \cite{BE85b}. In their work the star was 
defined as a 53 Per star with a tentative period of 1.3947\,d.\\
{\bf HD 57219 - $\nu$02 Pup.--} (55) Our analysis of this star's {\em Hipparcos} data shows little evidence
for variability, contrary to the suggestion of low-signal variability by
\cite{BCM92}. The spectral classification of the star is also a matter
of debate \citep[see][]{DES81}. \cite{RGC91} classify the star as B3 and  
He strong, which seems to be the classification most consistent with its
Str\"omgren colors.\\
{\bf HD 65575 - $\chi$ Car.--} (56) The {\em Hipparcos} light curves show no evidence for variability within
a limit of  3\,mmag.\\
{\bf HD 104841 - $\theta^2$ Cru.--} (57) Claimed to be an ultra-short period pulsator \citep{J79-552}, but not
confirmed. The {\em Hipparcos} photometry is consistent with a double-wave light
variation with a period of 3.4\,d.\\
{\bf HD 106490 - $\delta$ Cru.--} (58) Variability dubious, and if present, of long period (3.6\,d). 
{\em Hipparcos} photometry shows no variability above 3\,mmag.\\
{\bf HD 109668 - $\alpha$ Mus.--} (59) Variability dubious, and if present, of long period. {\em Hipparcos} 
photometry shows no variability above 2.5\,mmag. Radial velocity variable.\\
{\bf HD 122980 - $\chi$ Cen.--} (60) No short period light variations. The {\em Hipparcos} data 
indicate possible slower low-amplitude variability (this work). Radial 
velocity variable.\\
{\bf HD 130903 - V1018 Cen.--} (61) The {\em Hipparcos} data can be folded with a period of 1.65064 d to give 
a double-wave light curve. Although only few measurements are available, 
we suspect the star is a binary-induced variable.\\
{\bf HD 160762 $\iota$ Her.--} (62) Slowly pulsating B star with suspected, but unconfirmed 
shorter-period variations. Cannot be considered to be a $\beta$~Cephei 
star for the time being.\\
{\bf HD 188439 - V819 Cyg.--} (63) The {\em Hipparcos} period of this OB runaway star is 0.71373\,d, resulting in $Q=0.10$\,d.\\
{\bf HD 195556 - $\omega ^1$ Cyg.--} (64) The available data, including the {\em Hipparcos} photometry indicate 
several possible or unstable periods, all of which are however longer 
than 15\,hr.\\
{\bf HD 217811 - LN And.--} (65) Claimed to be an ultra-short period pulsator, but not confirmed. A 
3-day period explains the variations in the {\em Hipparcos} photometry.\\

\clearpage
\begin{deluxetable}{lll}
\tablecaption{Pulsation periods for stars from Table~1. The letter $\rm p$ after a given period denotes a photometric detection and $\rm s$ a spectroscopic one. Uncertainties of the periods are in the last digits.\label{tbl-4}}
\tablewidth{12cm}
\tablehead{
 \colhead{Identifier} & \colhead{Period [d]} & \colhead{Reference, (Note)}}
\startdata
       HD 886 & 0.1517502ps & 196\\
[3mm]  V909 Cas & 0.207p & 145\\
[3mm]  NGC\,663 4 & 0.194047p & 188\\
[3mm]  V611 Per & 0.1716946p & 019\\
[3mm]  V665 Per & 0.242342p & 146\\
 &  0.199545p & this work \\
 &  +more & \\
[3mm]  V614 Per & 0.1326359p & 019\\
[3mm]  NGC\,884 2246 & 0.184188p & 018\\
 &  0.170765p & \\
[3mm]  V595 Per & 0.31788p & 018\\
[3mm]  HD 16582 & 0.1611ps & s: 201, p: 202\\
[3mm]  HD 21803 & 0.201779ps & 203, 242\\
 &  0.198085ps & 204, 242\\
 &  0.227099p & 204, 242 \\
 &  +more & 242\\
[3mm]  HD 24760 & 0.1887s & 148; (66)\\
 &  0.1698s & \\
 &  0.1600s & \\
 &  0.1455s & \\
 &  0.13976s & s: 206, not found in 148\\
 &  0.1911s & s: 206, not found in 148 \\
[3mm]  HD 29248 & 0.1735126ps & 207\\
 &  0.1768681ps & \\
 &  0.1779337ps & \\
 &  0.1773937ps & \\
 &  0.126619ps & \\
 &  0.16015ps & \\
 &  0.15969ps & \\
 &  0.16074s & \\
 &  0.1389p & \\
[3mm]  HD 35411 & 0.133s & 208; (67)\\
[3mm]  HD 35715 & 0.0954s & 177\\
 &  0.0932s & \\
[3mm]  HD 44743 & 0.2512988ps & p: 011, s: 103, rv: 210\\
 &  0.25003ps & 011, 103, 211, rv: 210\\
 &  0.23904ps & 011, rv: 210, 212\\
[3mm]  HD 46328 & 0.2095754p & 094\\
[3mm]  HD 50707 & 0.18464ps & 094\\
 &  0.1932ps & 213\\
 &  0.1924p & \\
[3mm]  HD 52918 & 0.191207ps & 174; (68)\\
 &  0.204517ps & \\
[3mm]  HD 56014 & 0.0919p & p: 197 \\
[3mm]  HD 59864 & 0.238p: & 141; (69)\\
 &  0.243p: &\\
[3mm]  HD 61068 & 0.166385p & 094; (70)\\
 &  0.164921p &  \\
[3mm]  HD 64365 & 0.201584p& this work; (71)\\
& + more & \\
[3mm]  HD 64722 & 0.11541p & 214; (72)\\
 &  0.1168 or 0.1323p &\\
[3mm]  HD 71913 & 0.20578p & 058\\
[3mm]  HD 78616 & 0.21569ps & 094, 103; (73) \\
[3mm]  HD 80383 & 0.18316p & 003\\
 &  0.18647p & \\
 &  0.1847p & \\
[3mm]  HD 90288 & 0.10954p & 003 \\
 &  0.12024p & \\
 &  0.10344p & \\
 &  0.1295p & \\
[3mm]  HD 303068 & 0.1457p & 001; (74)\\
 &  0.1487p &\\
[3mm]  HD 303067 & 0.1684p & 001; similar situation as for HD 303068 \\
 &  0.1751p & \\
 &  0.1643p & \\
[3mm]  V403 Car & 0.251p & 001; similar situation as for HD 303068 \\
[3mm]  V412 Car & 0.114p: & 001 \\
[3mm]  V404 Car & 0.16p: & 001; (75) \\
[3mm]  V405 Car & 0.152p & 001\\
 &  0.158p & \\
 &  0.1841p: &094 \\
[3mm]  V378 Car & 0.1600p & 001 \\
 &  0.2070p & \\
 &  0.177p: & \\
[3mm]  V440 Car & 0.179p: & 140\\
[3mm]  V406 Car & 0.1756p & 001; similar situation as for HD 303068 \\
 &  0.1785p & \\
[3mm]  V380 Car & 0.2274p & 001 \\
[3mm]  V381 Car & 0.1773p & 001 \\
 &  0.1502p & 215 \\
 &  0.1397p & 215 \\
[3mm]  HD 109885 & 0.17054p & 003 \\
 &  0.16806p & \\
 &  0.1616p & \\
 &  0.1752p & \\
[3mm]  HD 111123 & 0.1911846ps & 216; (76)\\
 &  0.1678228ps & 217\\
 &  0.1827430ps & \\
[3mm]  BS Cru & 0.151p &004\\
 &  0.156p & \\
 &  0.163p & \\
 &  0.137p & \\
 &  0.157978p & 193 \\
[3mm]  NGC 4755\,113 & 0.233p & 002\\
[3mm]  NGC 4755\,405 & 0.125p & 004 \\
 &  0.128p & \\
[3mm]  CT Cru & 0.131p & 004\\
[3mm]  CV Cru & 0.179p & 004 \\
 &  0.128p & \\
[3mm]  CZ Cru & 0.159p & 004 \\
 &  0.108p & \\
 &  0.1386p & 001 \\
[3mm]  CX Cru & 0.182p & 004 \\
[3mm]  CY Cru & 0.159p & 002 \\
[3mm]  NGC 4755\,210 & 0.093p & 004 \\
[3mm]  BW Cru & 0.205p & 004 \\
 &  0.220p & \\
 &  0.190p & \\
 &  0.1623p & 193 \\
[3mm]  HD 112481 & 0.254537p & 094 \\
 &  0.259618p & \\
[3mm]  HD 116658 & 0.173787ps & 218; (77)\\
[3mm]  HD 118716 & 0.169608ps & 200, 232\\
 &  0.17696ps & 060, 232\\
 & 0.1617s & \\
 &  0.1356s & \\
 &  0.1308s & \\
[3mm]  HD 122451 & 0.153496s & 059\\
 & 0.155920s & (Balona's photometric period uncertain) \\
 & 0.153960s & \\
[3mm]  HD 126341 & 0.17736934ps & 219\\
[3mm]  HD 129056 & 0.25984663ps & 094 \\
 &  0.2368ps & 220\\
[3mm] HD 129557 & 0.1275504ps & 221\\
 &  0.142516p & 222\\
 &  0.134769p & \\
[3mm]  HD 129929 & 0.1547581p & 223\\
 &  0.1433013p & \\
 &  0.1550486p & \\
 &  0.1430527p & \\
 &  0.1517234p & \\
 &  0.1435509p & \\
[3mm]  HD 136298 & 0.198ps & 068, 200; (78)\\
[3mm]  HD 144470 & 0.067s & 092\\
[3mm]  HD 145794 & 0.15991p & 093; (79)\\
 &  0.1918p & \\
[3mm]  HD 147165 & 0.246829ps & 098\\
 &  0.239661ps & \\
[3mm]  HD 147985 & 0.132312ps & 102\\
 &  0.144930ps & 103\\
 &  0.156656ps & \\
[3mm]  V945 Sco & 0.06706p &100\\
[3mm]  V1032 Sco & 0.11928p & 015\\
 &  0.07699p & \\
 &  0.12040p & \\
[3mm]  V946 Sco & 0.09878p & 100 \\
 &  0.09544p & \\
 &  0.09071p & \\
 &  0.08550p & \\
 &  0.08302p & \\
[3mm] V964 Sco & 0.087846p & 015 \\
 &  0.067575p & \\
 &  0.055328p & \\
[3mm]  V947 Sco & 0.10788p & 100\\
 &  0.06096p & \\
[3mm]  V920 Sco & 0.10119p & 100 \\
 &  0.10765p & \\
 &  0.10389p & \\
 &  0.12137p & \\
 &  0.09114p & \\
[3mm]  HD 156327B & 0.146p & 185; (80)\\
 &  0.136p & \\
[3mm]  HD 156662 & 0.16890p & 102 \\
 &  0.18861p & \\
 &  0.16978p & \\
[3mm]  HD 157056 & 0.1405280ps & 106\\
 &  0.13722p & \\
 &  0.13569p & \\
 &  0.13391p & \\
 &  0.12877p & \\
 &  0.12699p & \\
 &  0.12542p & \\
[3mm]  HD 157485 & 0.2212p & 058 \\
 &  0.2240p & \\
[3mm]  HD 158926 & 0.2138272ps  & 108, see (24) \\
&+ more &\\
[3mm]  HD 160578 & 0.19983ps & 108, see (25) \\
[3mm]  HD 163472 & 0.13989010ps & 112, 242\\
 &  0.1466s & 113, 242\\
[3mm]  HD 164340 & 0.1529341p & 229\\
 &  0.1567948 & 229\\
[3mm]  HD 165174 & 0.303ps & 116, 242\\
[3mm]  HD 165812 & 0.1759p & 058 \\
 &  0.2180p & \\
[3mm]  HD 166540 & 0.23299p & 119\\
 &  0.22729p & \\
[3mm]  HD 180642 & 0.18225ps & 058, 242 \\
 & &  \\
[3mm]  NGC 6910\,18 & 0.156539p & 194\\
 &  0.162486p & \\
 &  0.148877p & \\
[3mm]  NGC 6910\,16 & 0.192198p & 194 \\
 &  0.171077p & \\
 &  0.239556p & \\
[3mm]  NGC 6910\,14 & 0.190396p & 194 \\
[3mm]  NGC 6910\,27 & 0.143010p & 194 \\
[3mm]  V2187 Cyg & 0.25388p & 176\\
[3mm]  HD 199140 & 0.20104444ps & 122, 224\\
[3mm]  HD 203664 & 0.16587ps & 058, 242\\
 &  +more & 242\\
[3mm]  HD 205021 & 0.1904870ps & 126\\
 &  0.2031s & 225\\
 &  0.1967s & \\
 &  0.1859s & \\
 &  0.18460s & \\
[3mm]  NGC 7235\,8 & 0.202890p & 186\\
 &  0.177898p & \\
[3mm]  HN Aqr & 0.15231ps & 129, 130 \\
[3mm]  HD 214993 & 0.23583ps & 091\\
 &  0.19738ps & 226\\
 &  0.19309ps & \\
 &  0.1917p & \\
 &  0.1884p & \\
 &  0.18747ps & \\
 &  0.18215ps & \\
 &  0.1711p & \\
 &  0.1350p & \\
[3mm]  HD 216916 & 0.1691670ps & 227\\
 &  0.1708555ps & 240\\
 &  0.1817325ps & \\
 &  0.1816843p &
\enddata
\end{deluxetable}

\clearpage
\subsection{Notes on individual frequencies}

{\bf HD 24760 - $\epsilon$ Per.--} (66) A period at 0.0945\,d was detected by \cite{SFP87} as well as
\cite{GKJ99}. We assume that this is a harmonic of the main pulsation
mode at $P$=0.1887\,d.\\
{\bf HD 35411 - $\eta$ Ori.--} (67) A period at 0.43208\,d was found by \cite{WL88} in their photometric
data. It is doubtful if it originates from pulsation.\\
{\bf HD 52918 - 19 Mon.--} (68) A period at 5.88\,d was also detected \citep{BJM02}, which is too long for $\beta$~Cephei-type pulsation.\\
{\bf HD 59864 - V350 Pup.--} (69) \cite{SJ90} demonstrated the multiperiodicity of this star, but could 
not give unambiguous period determinations due to aliasing. The choice of 
the primary frequency also affects all others. The periods we list are the 
most likely ones from the work by \cite{SJ90}.\\
{\bf HD 61068 - PT Pup.--} (70) \cite{H92} reports two frequencies that we list in Table~4. This
author however noted aliasing problems in his period determinations.  
Amplitude variability also seems present. In addition, the {\em Hipparcos} data
for this star \citep{KE02} do not confirm the periodicities listed
by \cite{H92}. More observations of this star are clearly necessary to
determine the periodic content of its variability properly. \\ 
{\bf HD 64365 - QU Pup.--} (71) Frequency analyses by \cite{SJ80} and \cite{H92} had periods of
0.1678\,d and 0.1927\,d in common. However, our analysis of the {\em Hipparcos}
photometry of that star resulted in a 0.2016\,d period, which is the 1 c/d
alias of the 0.1678\,d period given by the previous authors. As we suspect
that the prewhitening of this erroneous period from single-site data had
generated spurious secondary signals, we only list the frequency found in
the {\em Hipparcos} data. We do point out that the star is multiperiodic in any 
case.\\
{\bf HD 64722 - V372 Car.--} (72) Aliasing mistake by \cite{H92}, solved by re-analysis of {\em Hipparcos} 
data, this work.\\
{\bf HD 78616 - KK Vel.--} (73) There may be another independent pulsation mode at half the period 
we listed.\\
{\bf HD 303068.--} (74) Different authors list up to 5 different frequencies
\cite[see][]{E86,H92}, of which only two are in common in the different 
studies.\\
{\bf V404 Car.--} (75) Additional periods of 0.1742\,d and 0.1506\,d are listed by \cite{E86}, but they were not confirmed by other work.\\
{\bf HD 111123 - $\beta$ Cru.--} (76) More frequencies are possibly present, but we are unsure whether they 
originate from pulsation \citep{CAB02}.\\
{\bf HD 116658 - $\alpha$ Vir.--} (77) The only periodicity that we regard convincing in the
analyses of this star is 0.1738\,d. \cite{Sm85} found a number of
additional signals in his line-profile analysis. We support the suggestion by
\cite{AD03} that more spectroscopic data have to be analyzed
before a definite conclusion about the presence of the additional
periodicities can be made.\\
{\bf HD 136298 - $\delta$ Lup.--} (78) Photometric period likely an alias of the spectroscopic one quoted.\\
{\bf HD 145794 - V349 Nor.--} (79) The value of the second period of this star is uncertain due to 
aliasing \citep{WH89}.\\
{\bf HD 156327B - V1035 Sco.--} (80) The V amplitude is 35\,mmag; spectroscopic variability was detected, but
no period could be determined.\\

\section{List of references and according numbers in the text}

\hspace{-4.3mm}$[$001$]$ \cite{BDP97} \\
$[$002$]$ \cite{BK94} \\
$[$003$]$ \cite{HSV03} \\
$[$004$]$ \cite{SHH02} \\
$[$005$]$ \cite{CDP94} \\
$[$006$]$ \cite{B75} \\
$[$007$]$ \cite{SIF03} \\
$[$008$]$ \cite{LA78} \\
$[$009$]$ \cite{DFM95} \\
$[$010$]$ \cite{E67} \\
$[$011$]$ \cite{S73-161}\\
$[$012$]$ \cite{KH75} \\
$[$013$]$ \cite{F63} \\
$[$014$]$ \cite{W91} \\
$[$015$]$ \cite{BE85a} \\
$[$016$]$ \cite{CR98} \\
$[$017$]$ \cite{KE02} \\
$[$018$]$ \cite{KP97} \\
$[$019$]$ \cite{KPK99} \\
$[$020$]$ \cite{McN85} \\
$[$021$]$ \cite{B90} \\
$[$022$]$ \cite{BCM92} \\
$[$023$]$ \cite{KSD99} \\
$[$024$]$ \cite{J93aaps} \\
$[$025$]$ \cite{FGB91} \\
$[$026$]$ \cite{JS77} \\
$[$027$]$ \cite{M85} \\
$[$028$]$ \cite{SNC99} \\
$[$029$]$ \cite{ACP99} \\
$[$030$]$ \cite{AB96} \\
$[$031$]$ \cite{HOD76} \\
$[$032$]$ \cite{Sxx} \\
$[$033$]$ \cite{Ja78} \\
$[$034$]$ \cite{BL95} \\
$[$035$]$ \cite{GM01} \\
$[$036$]$ \cite{ASK01} \\
$[$037$]$ \cite{PHK02} \\
$[$038$]$ \cite{PHB97} \\
$[$039$]$ \cite{H67a} \\
$[$040$]$ \cite{PM72} \\
$[$041$]$ \cite{J60} \\
$[$042$]$ \cite{DAG92} \\
$[$043$]$ \cite{B77} \\
$[$044$]$ \cite{BE81} \\
$[$045$]$ \cite{VWP74} \\
$[$046$]$ \cite{E79-IBVS} \\
$[$047$]$ \cite{S81} \\
$[$048$]$ \cite{D70} \\
$[$049$]$ \cite{F58} \\
$[$050$]$ \cite{HKB74} \\
$[$051$]$ \cite{AD03} \\
$[$052$]$ \cite{GM68} \\
$[$053$]$ \cite{DK84} \\
$[$054$]$ \cite{PFL76} \\
$[$055$]$ \cite{H70} \\
$[$056$]$ \cite{S73-162} \\
$[$057$]$ \cite{W53} \\
$[$058$]$ \cite{A00} \\
$[$059$]$ \cite{AAU02} \\
$[$060$]$ \cite{S99} \\
$[$061$]$ \cite{HHH97} \\
$[$062$]$ \cite{ADH04} \\
$[$063$]$ \cite{RBA99} \\
$[$064$]$ \cite{BP70} \\
$[$065$]$ \cite{H71} \\
$[$066$]$ \cite{ATD03} \\
$[$067$]$ \cite{KVS00} \\
$[$068$]$ \cite{LP88} \\
$[$069$]$ \cite{S70} \\
$[$070$]$ \cite{LMC01} \\
$[$071$]$ \cite{KKA81} \\
$[$072$]$ \cite{MM79} \\
$[$073$]$ \cite{I75}  \\
$[$074$]$ \cite{P84} \\
$[$075$]$ \cite{MB91} \\
$[$076$]$ \cite{LMM87} \\
$[$077$]$ \cite{KKP71} \\
$[$078$]$ \cite{PA00} \\
$[$079$]$ \cite{J79-552} \\
$[$080$]$ \cite{J79-1042} \\
$[$081$]$ \cite{vH73} \\
$[$082$]$ \cite{WR83b} \\
$[$083$]$ \cite{RS77} \\
$[$084$]$ \cite{GBS87} \\
$[$085$]$ \cite{LC98} \\
$[$086$]$ \cite{Y01} \\
$[$087$]$ \cite{S83a} \\
$[$088$]$ \cite{BE85b} \\
$[$089$]$ \cite{S78} \\
$[$090$]$ \cite{JKM03} \\
$[$091$]$ \cite{Het04} \\
$[$092$]$ \cite{TS98a} \\
$[$093$]$ \cite{WH89} \\
$[$094$]$ \cite{H92} \\
$[$095$]$ \cite{HWS94} \\
$[$096$]$ \cite{GLM84} \\
$[$097$]$ \cite{JS84} \\
$[$098$]$ \cite{CV92} \\
$[$099$]$ \cite{B83} \\
$[$100$]$ \cite{BS83} \\
$[$101$]$ \cite{S79} \\
$[$102$]$ \cite{WC85} \\
$[$103$]$ \cite{AWP94} \\
$[$104$]$ \cite{H22} \\
$[$105$]$ \cite{B71} \\
$[$106$]$ \cite{HSM03} \\
$[$107$]$ \cite{SL72} \\
$[$108$]$ \cite{LS75} \\
$[$109$]$ \cite{MAW97} \\
$[$110$]$ \cite{UAD01} \\
$[$111$]$ \cite{Je72} \\
$[$112$]$ \cite{KS84} \\
$[$113$]$ \cite{NHF03} \\
$[$114$]$ \cite{L59} \\
$[$115$]$ \cite{Je75} \\
$[$116$]$ \cite{CBM89} \\
$[$117$]$ \cite{CK93} \\
$[$118$]$ \cite{WAK98} \\
$[$119$]$ \cite{WAW91} \\
$[$120$]$ \cite{PP31} \\
$[$121$]$ \cite{LS87} \\
$[$122$]$ \cite{AMV95} \\
$[$123$]$ \cite{HKK94} \\
$[$124$]$ \cite{F06} \\
$[$125$]$ \cite{PB92} \\
$[$126$]$ \cite{TAM97} \\
$[$127$]$ \cite{SA00} \\
$[$128$]$ \cite{WR88} \\
$[$129$]$ \cite{KW90} \\
$[$130$]$ \cite{DKK98} \\
$[$131$]$ \cite{Y15} \\
$[$132$]$ \cite{J78} \\
$[$133$]$ \cite{A96} \\
$[$134$]$ \cite{DJ99} \\
$[$135$]$ \cite{W51} \\
$[$137$]$ \cite{DJ96} \\
$[$138$]$ \cite{TAD03} \\
$[$139$]$ \cite{K01} \\
$[$140$]$ \cite{B94} \\
$[$141$]$ \cite{SJ90} \\
$[$142$]$ \cite{SJ93} \\
$[$143$]$ \cite{E86} \\
$[$144$]$ \cite{B95a} \\
$[$145$]$ \cite{RDC00} \\
$[$146$]$ \cite{G00} \\
$[$147$]$ \cite{S68} \\
$[$148$]$ \cite{CTA00} \\
$[$149$]$ \cite{ALG02} \\
$[$150$]$ \cite{SS97} \\
$[$151$]$ \cite{E79-ESO} \\
$[$152$]$ \cite{McS03} \\
$[$153$]$ \cite{A01} \\
$[$154$]$ \cite{RGM99} \\
$[$155$]$ \cite{CSV98} \\
$[$156$]$ \cite{MAB01} \\
$[$157$]$ \cite{JJH80} \\
$[$158$]$ \cite{M94} \\
$[$159$]$ \cite{WR83a} \\
$[$160$]$ \cite{B82} \\
$[$161$]$ \cite{ESA97} \\
$[$162$]$ \cite{MW95} \\
$[$163$]$ \cite{CML00} \\
$[$164$]$ \cite{GD82} \\
$[$165$]$ \cite{BE86} \\
$[$166$]$ \cite{DGG84} \\
$[$167$]$ \cite{DAG85} \\
$[$168$]$ \cite{HJI61} \\
$[$169$]$ \cite{MWH91} \\
$[$170$]$ \cite{SFM83} \\
$[$171$]$ \cite{PL77} \\
$[$172$]$ \cite{DES81}\\
$[$173$]$ \cite{H82} \\
$[$174$]$ \cite{BJM02} \\
$[$175$]$ \cite{WR85} \\
$[$176$]$ \cite{PK98} \\
$[$177$]$ \cite{TAS01} \\
$[$178$]$ \cite{SL83} \\
$[$179$]$ \cite{vH73Med} \\
$[$180$]$ \cite{KL96} \\
$[$181$]$ \cite{K94} \\
$[$182$]$ \cite{B95b}\\ 
$[$183$]$ \cite{JSK88}\\
$[$184$]$ \cite{Sm81}\\
$[$185$]$ \cite{PVG02}\\
$[$186$]$ \cite{PJK97}\\
$[$187$]$ \cite{P97}\\
$[$188$]$ \cite{PKK01}\\
$[$189$]$ \cite{BMD94}\\
$[$190$]$ \cite{KPK02}\\
$[$191$]$ \cite{BBG91}\\
$[$192$]$ \cite{BC93}\\
$[$193$]$ \cite{K93}\\
$[$194$]$ \cite{KKP04}\\
$[$195$]$ \cite{HSJ04}\\
$[$196$]$ \cite{VCM85}\\
$[$197$]$ \cite{BKr94}\\
$[$198$]$ \cite{ST02}\\
$[$199$]$ \cite{C82}\\
$[$200$]$ \cite{S72}\\
$[$201$]$ \cite{CS80}\\
$[$202$]$ \cite{CN97}\\
$[$203$]$ \cite{JJR81}\\
$[$204$]$ \cite{SZ59}\\
$[$205$]$ \cite{SFP87}\\
$[$206$]$ \cite{GKJ99}\\
$[$207$]$ \cite{DTB04}\\
$[$208$]$ \cite{MAW96}\\
$[$209$]$ \cite{WL88}\\
$[$210$]$ \cite{S50}\\
$[$211$]$ \cite{BBL96}\\
$[$212$]$ \cite{K80}\\
$[$213$]$ \cite{LSS56}\\
$[$214$]$ \cite{SJ80}\\
$[$215$]$ \cite{FHS03}\\
$[$216$]$ \cite{CAB02}\\
$[$217$]$ \cite{ACC98}\\
$[$218$]$ \cite{L78}\\
$[$219$]$ \cite{C87}\\
$[$220$]$ \cite{MAP94}\\
$[$221$]$ \cite{LS85}\\
$[$222$]$ \cite{SJ83}\\
$[$223$]$ \cite{AWD04}\\
$[$224$]$ \cite{SPL93}\\
$[$225$]$ \cite{SK54}\\
$[$226$]$ \cite{MAG94}\\
$[$227$]$ \cite{LHA01}\\
$[$228$]$ \cite{MCN04}\\
$[$229$]$ \cite{MP04}\\
$[$230$]$ \cite{STD02}\\
$[$231$]$ \cite{TUI02}\\
$[$232$]$ \cite{STA04}\\
$[$233$]$ \cite{LiH98}\\
$[$234$]$ \cite{GFB01}\\
$[$235$]$ \cite{Sh85}\\
$[$236$]$ \cite{BFM89}\\
$[$237$]$ \cite{M85}\\
$[$238$]$ \cite{PT77}\\
$[$239$]$ \cite{vH63}\\
$[$240$]$ \cite{JP99}\\
$[$241$]$ \cite{GHS77}\\
$[$242$]$ \cite{DDU04}\\
$[$243$]$ \cite{S88}\\

\acknowledgments

During most of this work, AS was a European Space Agency Post-doctoral Research Fellow. GH acknowledges support from the Austrian Fonds zur F\"orderung der wissenschaftlichen Forschung under grant R12-N02.

Several people have considerably helped the authors during the course of this work. First and foremost, we are indebted to Peter De Cat for his generous support of this work and for his valuable comments that improved this paper. One of the starting points for this work was his online catalog of $\beta$~Cephei stars.

We are grateful to Alosha Pamyatnykh for sending us his compilation of $\beta$~Cephei stars, for supplying theoretical instability strip boundaries and for permission to use the Warsaw-New Jersey code. We also thank Andrzej Pigulski and Katrien Uytterhoeven for making some results available to us before publication, as well as Chris Sterken and Lars Freyhammer for helpful comments on some stars. Finally, we wish to thank Mike Jerzykiewicz, Conny Aerts, and (again) Alosha Pamyatnykh, Andrzej Pigulski, and Peter De Cat for their valuable comments and suggestions on a draft version of this paper.

This research has made use of the SIMBAD database operated at CDS Strasbourg, France. In this work we made use of The General Catalog of Photometric Data (GCPD) II by \cite{MMH97}.

\clearpage

\end{document}